\documentclass[preprint,pre,longbibliography]{revtex4-1}
\usepackage{amsfonts,amssymb,amsbsy}
\usepackage{amsmath}
\usepackage{latexsym}
\usepackage{graphicx}
\usepackage{lineno}
\usepackage{color}
\usepackage{natbib}

\newcommand{\la}{\langle}
\newcommand{\ra}{\rangle}
\newcommand{\eps}{\epsilon}
\newcommand{\nb}{\boldsymbol{\nabla}^\bot}

\def\bx{\bx}
\def\bxi{\boldsymbol{\xi}}
\def\Re{\mathop{\mathrm{Re}}}
\def\bx{\boldsymbol{x}}
\def\by{\boldsymbol{y}}
\def\bk{\boldsymbol{k}}
\def\bp{\boldsymbol{p}}
\def\bv{\boldsymbol{v}}
\def\bl{\boldsymbol{l}}
\def\e{\mathrm{e}}
\def\d{\mathrm{d}}
\def\i{\mathrm{i}}
\def\lw{\ell_\mathrm{w}}
\def\lf{\ell_\mathrm{f}}
\def\lc{\ell_\mathrm{c}}
\def\kc{k_\mathrm{c}}

\newcommand*\patchAmsMathEnvironmentForLineno[1]{%
  \expandafter\let\csname old#1\expandafter\endcsname\csname #1\endcsname
  \expandafter\let\csname oldend#1\expandafter\endcsname\csname end#1\endcsname
  \renewenvironment{#1}%
     {\linenomath\csname old#1\endcsname}%
     {\csname oldend#1\endcsname\endlinenomath}}%
\newcommand*\patchBothAmsMathEnvironmentsForLineno[1]{%
  \patchAmsMathEnvironmentForLineno{#1}%
  \patchAmsMathEnvironmentForLineno{#1*}}%
\AtBeginDocument{%
\patchBothAmsMathEnvironmentsForLineno{equation}%
\patchBothAmsMathEnvironmentsForLineno{align}%
\patchBothAmsMathEnvironmentsForLineno{flalign}%
\patchBothAmsMathEnvironmentsForLineno{alignat}%
\patchBothAmsMathEnvironmentsForLineno{gather}%
\patchBothAmsMathEnvironmentsForLineno{multline}%
}

\begin{document}

\title{Near-inertial wave scattering by random flows}
\author{Eric Danioux}
\affiliation{ School of Mathematics and Maxwell Institute for Mathematical Sciences,
University of Edinburgh, Edinburgh EH9 3JZ, UK}
\author{Jacques Vanneste}
\email[Corresponding author: ]{J.Vanneste@ed.ac.uk}
\affiliation{ School of Mathematics and Maxwell Institute for Mathematical Sciences,
University of Edinburgh, Edinburgh EH9 3JZ, UK}

\date{\today}

\begin{abstract}
The impact of a turbulent flow on wind-driven oceanic near-inertial waves is examined using a linearised shallow-water model of the mixed layer. Modelling the flow as a homogeneous and stationary random process with spatial scales comparable to the wavelengths, we derive a transport (or kinetic) equation governing wave-energy transfers in both physical and spectral spaces. This equation describes the scattering of the waves by the flow which results in a redistribution of energy between waves with the same frequency (or, equivalently, with the same wavenumber) and, for isotropic flows, in the isotropisation of the wave field. The time scales for the scattering and isotropisation are obtained explicitly and found to be of the order of tens of days for typical oceanic parameters. The predictions inferred from the transport equation are confirmed by a series of numerical simulations. 

Two situations in which near-inertial waves are strongly influenced by flow scattering are investigated through dedicated nonlinear shallow-water simulations. In the first, a wavepacket propagating equatorwards  as a result from the $\beta$-effect is shown to be slowed down and dispersed both zonally and meridionally by scattering. In the second, waves generated by moving cyclones are shown to be strongly disturbed by scattering, leading again to an increased dispersion. 

\end{abstract}

\maketitle

\section{Introduction}

Near-inertial waves (NIWs), that is, internal waves with frequencies close to the local Coriolis frequency, are a major source of variability in the ocean where they play an important dynamical role (see \citep{Alford2016} for a recent review of NIW observations and models).  They are mainly generated at the surface by wind-stress forcing, although other generation mechanisms exist including geostrophic adjustment \citep{Blumen72}, flow-topography interaction \citep{Gill82}, momentum deposition \citep{Lott03, Nikurashin10}, and spontaneous generation \citep{Danioux12,Alford2013}. Wind-generated NIWs have small vertical scales, leading to strong vertical shear, instabilities and, as a result, surface mixing. Because of this and because of the vertical motion they induce, they have a strong impact on biological production \citep{Zhang2014}. NIWs  also propagate to the deep ocean \citep{Alford2012} where they eventually dissipate and induce diapycnal mixing, potentially sustaining the meridional overturning circulation; the extent of their contribution to the latter is still a matter of debate \citep{Furuichi2008}. There is also mounting evidence of a direct energetic impact of NIWs on the mesoscale flow \citep{Whitt2013,Whitt2015,Xie2015}.


Regions of NIW-generation usually coincide with regions of strong geostrophic turbulence (see for example  Fig.\ 1 in  \citep{Zhai05}). This has  motivated numerous studies of the impact of a flow on the propagation of NIWs \citep{Mooers75a,Mooers75b,Olbers81a,Kunze85,YBJ,Klein04,Danioux08a,Whitt2013}. 
Because of the different atmospheric and oceanic Rossby radii, the horizontal scales of the wind patterns and hence of the NIW forcing are often much larger than scales of the oceanic flow. Consequently, many of the aforementioned studies take the NIWs as initially homogeneous. However, as shown by \citet{Dasaro85} and confirmed by \citet{Alford03a} and \citet{Silverthorne2009}, most of the NIW energy input to the ocean is due to a few intense winter storms or cyclones, with horizontal scales of hundreds of kilometers. These storms move horizontally, exciting NIWs at scales that depend on their speed and can be  
comparable to oceanic scales, then invalidating the assumption of homogeneous initial condition. There is, then, a need to understand how the flow affects NIWs generated at scales comparable to flow scales. The primary effect is the scattering of the NIWs, leading to a redistribution of their energy in wavenumber space. Describing and quantifying this scattering are the aims of the present paper.



We adopt the framework developed by \citet{Ryzhik96} for a broad class of scattering problems (see also Ref.\ \citenum{Powell05}). This describes asymptotically the propagation of a spectrum of waves in a medium -- the flow in our case --  that varies randomly over spatial scales similar to the wavelength. The approach represents the wave field by its Wigner transform to account for large-scale spatial modulations of the wave spectrum. Taking the model of NIW propagation derived by \citet[][referred below as YBJ]{YBJ} as our starting point, we derive the transport equation satisfied by the associated Wigner transform and examine several predictions of this equation. The principal one is the spreading of the NIW energy spectrum along lines of constant frequency (corresponding to constant horizontal wavenumber for NIWs), leading to a relaxation towards an isotropic spectrum when the flow is itself isotropic. We estimate the timescale for this relaxation, and find it to be of the order of a few weeks, depending on parameters including the flow strength and  correlation length. We test our analytic predictions against numerical simulations of a reduced-gravity shallow-water version of the YBJ model. 

The importance of scattering for NIWs and the relevance of our theoretical conclusions are illustrated by two applications which we discuss on the basis of simulations of the full nonlinear reduced-gravity shallow-water model. This model introduces more realism through effects not included in the YBJ and scattering theories. The first application considers the interplay between  scattering and the $\beta$-effect; it shows that this interplay leads a slowdown of the equatorward propagation of NIW wavepackets. The second application considers NIWs generated by a moving cyclone and identifies a speedup of lateral dispersion as one of the main impacts of the flow.

The plan of the paper is as follows. In section \ref{sec:review}, we briefly review various approaches to the study of NIW propagation in background flows. Section \ref{sec:NIW_scat} introduces the transport equation obeyed by the Wigner function associated with NIWs. Because of its mathematical complexity, the derivation of this equation is relegated to Appendix \ref{app:deriv}. Section \ref{sec:NIW_scat} also shows how the isotropisation of the wave field results from the scattering, estimates the relevant timescales as a function of the flow parameters, and verifies the analytic predictions numerically. The two applications 
are presented in section \ref{sec:storm}. The paper concludes with a discussion in section \ref{sec:discussion}. 

\section{NIW propagation in flows} \label{sec:review}

Many studies of the propagation of oceanic internal waves in heterogeneous flows rely on the 
Wentzel--Kramers--Brillouin (WKB) approximation \citep{Mooers75a,Mooers75b, Kunze85} which assumes that the length scale $\lw$  of the waves is much smaller than the scale $\lf$ of the background flow. Wind-driven NIWs, however, typically span a broad range of scales, from the large atmospheric scales at which they are generated to the much smaller scales they reach as a result of advection and refraction by the flow.  
To account for this,  \citet{YBJ} proposed a model for the propagation of NIWs in a geostrophic flow without any assumptions on their relative scale $\lw/\lf$. Using a multiple time-scale approach (where the slow time scale is related to the small NIW-Burger number), they derived an equation for the subinertial amplitude $M$ from which the NIW
complex velocity is deduced as $u+\i v=M \e^{-\i f_0t}$.
With the additional assumption of a barotropic geostrophic flow with streamfunction $\psi(x,y,t)$, the amplitude of each vertical mode obeys the YBJ equation
\begin{equation}\label{eq:YBJ}
\partial_t M + \nb\psi\cdot \boldsymbol{\nabla} M - \i\frac{h}{2}\Delta M + \i\frac{\Delta \psi}{2} M=0,
\end{equation}
where $h=f_0r_n^2$ depends on the NIW-vertical mode $n$ through its Rossby radius of deformation $r_n$, and $\nb=(-\partial_y,\partial_x)$. This equation, which holds provided that
\begin{equation}
\mathrm{Bu} = {r_n^2}/{\lw^2} \ll 1,
\end{equation}
exhibits the three physical mechanisms influencing the evolution of NIWs, namely advection, dispersion and refraction. It also applies to NIWs concentrated in a homogeneous layer capping an abyssal ocean where the only motion is the barotropic flow, as described by a reduced-gravity shallow-water model; in this case, $h=g'H/f_0$ where $g'$ is the reduced gravity and $H$ is the depth of the top layer, and $\mathrm{Bu} = g' H/(f^2 \lw^2)$ (see \citep{Danioux15}). This is the interpretation that we take in this paper, focussing on applications to mixed-layer NIWs.  


Most of the work on (\ref{eq:YBJ}) considers the case of homogeneous initial NIWs, that is, infinite initial $\lw$ or, more broadly, $\lw \gg \lf$  \citep[e.g.][]{YBJ,Balmforth98,Klein04}. Finite length scales appear as a result of the interactions with the flow, leading to a ratio $\lw/\lf$ that depends on the value of $h/\Psi$. In the long-time limit, this ratio becomes $O(\sqrt{h/\Psi})$ in the strong dispersion regime $h \gg \Psi$, $O(1)$ in the intermediate regime $h=O(\Psi)$, and $h/\Psi$ in the strong trapping regime $h \ll \Psi$ \citep{Danioux15}. The opposite limit, where $\lw \ll \lf$ initially can be tackled using the WKB approximation. It is clear from (\ref{eq:YBJ}) that the refraction term becomes much smaller than the advection term in this limit, except for some very specific flow configurations, as noted by \citet{Olbers81a}.

\begin{table}
\begin{center}
\begin{tabular}{ccc}
 & spatial scale & time scale \\
\hline
wave phase & $\ell$ & $\ell^2/h$ \\
wave envelope & $L =  \ell/\eps$ & $L \ell / h = \eps L^2/h$ \\
flow &  $\ell$ & $L \ell / h = \eps L^2/h$ \\
\hline
\end{tabular}
\caption{Scaling assumptions: the wave time scales are deduced from the spatial scales using the dispersion relation $\omega = h |\bk|^2/2$ for the frequency shift relative to the inertial frequency $f_0$. The flow amplitude is determined by the scaling $\Psi=O(\eps^{1/2} h)$ of the streamfunction.}
\label{table:sc}
\end{center}
\end{table}

The case where $\lw \sim \lf$ initially remains largely unexplored despite its relevance to oceanic situations (see section \ref{sec:storm}). It is the focus of this paper. We make analytical progress by assuming that the NIW field consists of a spectrum of waves whose phase varies on the flow scale, thus taking $\lw=\lf=:\ell$ from now on. Inhomogeneities in the wave field are accounted for by considering an
amplitude that varies on a much larger length scale $L=\ell/\eps$, where $\epsilon\ll 1$. The flow is taken to be relatively weak, specifically such that $\Psi/h=O(\epsilon^{1/2})$, and modelled by a homogeneous and stationary random process in space and time. The scaling assumptions are further discussed in Appendix  \ref{app:deriv} and summarised in Table \ref{table:sc}.
We next derive the transport equation that governs the dynamics of the wave field in this setup. 

\section{NIW scattering} \label{sec:NIW_scat}


We adopt the approach of \citet{Ryzhik96}, formulated in terms of the Wigner transform which we now introduce.

\subsection{Wigner transform}

The Wigner transform of a function $M(\bx,t)$ rapidly decaying at infinity is defined as
\begin{equation}\label{eq:Wig}
W(\bx,\bk,t)=\frac{1}{4\pi^2}\int_{\mathbb{R}^2} \e^{\i\bk\cdot\by}M(\bx-{\by}/{2},t)M^*(\bx+{\by}/{2},t) \, \d\by,
\end{equation}
where $^*$ denotes the complex conjugate. Here, both $\bx$ and $\bk$ are two-dimensional (horizontal) vectors. It is easy to show that $W(\bx,\bk,t)$ is real and that its integral over wavevector space is
\begin{equation}\label{eq:int_Wig}
\int_{\mathbb{R}^2} W(\bx,\bk,t) \, \d\bk=|M(\bx,t)|^2.
\end{equation}
In the context of the YBJ equation, this quantity is twice the local NIW kinetic energy.
For a wavepacket solution
\begin{equation}\label{eq:wave_solution}
M(\bx,t)=A(\bx,t) \e^{\i\bk_0 \cdot\bx/\eps},
\end{equation}
where $A$ is a smooth function of $\bx$ and $|\bk_0|=O(1)$, the Wigner transform tends to
$$W(\bx,\bk,t)\rightarrow|A(\bx,t)|^2\delta(\bk-\bk_0/\eps)$$
as the scale-separation parameter $\eps\rightarrow 0$.
Hence, the Wigner transform can be  intuitively thought  of  as a wavenumber-resolving energy density in the scale-separation regime, and therefore bears similarities with wavelet transforms.

\subsection{Transport equation}

We assume that the NIW field consists of a spectrum of wavepackets of the form (\ref{eq:wave_solution}), with phases that vary on the  flow scale and amplitudes that vary on a much larger scale. In other words, the waves fluctuate on the flow scale, but are modulated over a larger envelope scale, with the scale separation measured by $\eps\ll 1$. This assumption implies that the scaled Wigner transform
\begin{equation}
W^\eps(\bx,\bk,t) = \eps^{-2} W(\bx,\bk/\eps,t),
\end{equation}
where $\bx$ and $\bk$ are non-dimensionalised using the large envelope scale $L=\eps^{-1} \ell$, tends to a finite limit $W^0$ as $\eps \to 0$. The flow, represented by a random streamfunction $\psi(\bx,t)$ with stationary and homogeneous statistics, has an amplitude $\Psi$ that satisfies the scaling $\Psi/h = O(\eps^{1/2})$ (see Appendix \ref{app:deriv}). This ensures that effects associated with the flow are smaller than dispersion effects in Eq.\ (\ref{eq:YBJ}). In stronger flows, waves simply do not propagate while weaker flows do not affect wave propagation on realistic timescales.

Under these assumptions, the leading-order Wigner transform $W^0$ of $M$, denoted below simply by $W$, satisfies the transport equation 
\begin{equation}\label{eq:transport}
\partial_tW+h\bk\cdot\boldsymbol{\nabla}_{\bx} W=\mathcal{L}W-\Sigma(\bk)W,
\end{equation}
where 
\begin{equation}\label{eq:LW}
\mathcal{L}W=\int_{\mathbb{R}^2}\sigma(\bk,\bp)W(\bx,\bp,t) \, \d\bp 
\end{equation}
and
\begin{equation}\label{eq:tot_scat}
\Sigma(\bk)=\int_{\mathbb{R}^2}\sigma(\bk,\bp) \, \d\bp
\end{equation}
(see Appendix \ref{app:deriv} for the derivation). Here,
\begin{equation}\label{eq:scat}
\sigma(\bk,\bp)=\frac{4\pi}{h}\left(|\bk\times\bp|^2+\frac{|\bk-\bp|^4}{4}\right)\hat{R}(\bp-\bk)\delta(\bk^2-\bp^2),
\end{equation}
where $\hat{R}(\bp)$ is the power spectrum of the streamfunction (that is, the Fourier transform of its  covariance, see (\ref{eq:cov})).
It is independent of time because of the assumed stationarity of the flow.
Under the additional assumption of isotropy, which we will make, it depends only on the magnitude $|\bp|$ of $\bp$.

The function $\sigma(\bk,\bp)$ is interpreted as a differential scattering cross-section, representing the rate at which energy at wavevector $\bp$ is converted to energy at wavevector $\bk$. In general, it is a function of space and time but this dependence drops here because of the homogeneity and stationarity of the flow. The total scattering cross-section $\Sigma(\bk)$ represents the rate at which energy at wavevector $\bk$ is transferred to all other wavevectors.
Importantly, the particular form of (\ref{eq:scat}) shows that energy transfers are restricted to wavevectors with the same magnitude:
\begin{equation}\label{eq:prop}
\sigma(\bk,\bp)=0 \ \ \text{if} \ \ |\bk|\neq|\bp|.
\end{equation} 
This is the particularisation to NIWs, with dispersion relation $\omega(\bk)=h |\bk|^2/2$ (see (\ref{eq:YBJ})), of the general restriction of energy transfers between waves of equal frequencies: $\omega(\bk)=\omega(\bp)$.
The two terms in (\ref{eq:scat}) stem from the advection and refraction terms, respectively. Wave propagation is governed by the second term on the left-hand side of equation (\ref{eq:transport}), which can be identified as  advection by the group velocity $\nabla_{\bk} \omega(\bk) = h \bk$ in a WKB context.

Integration of (\ref{eq:transport}) with respect to $\bk$ and  use of (\ref{eq:int_Wig}) and of the property $\sigma(\bk,\bp)=\sigma(\bp,\bk)$ give the leading-order energy conservation equation 
\begin{equation}
\partial_t\frac{1}{2}|M|^2+\boldsymbol{\nabla}_{\bx}\cdot\mathbf{F}=0,
\end{equation} 
where
$$\mathbf{F}=\frac{h}{2}\int_{\mathbb{R}^2} \bk W(\bx,\bk) \, \d\bk=\frac{ih}{4}(M\boldsymbol{\nabla}_{\bx}
 M^*-M^*\boldsymbol{\nabla}_xM)$$
is the  NIW kinetic energy flux \citep{YBJ,Danioux15}.


\subsection{Isotropisation}
\subsubsection{General properties}
In view of (\ref{eq:prop}), it is convenient to use polar coordinates for the wavevector. This simplifies (\ref{eq:LW})  into the single integral
\begin{equation}\label{eq:L_theta}
\mathcal{L}W(\bx,|\bk|,\theta,t)=\int_{-\pi}^{\pi}\sigma'(|\bk|,\theta')W(\bx,|\bk|,\theta+\theta',t) \, \d\theta',
\end{equation}
where $\theta$ is the orientation of $\bk$, $\theta'$ is the angle between $\bk$ and $\bp$ in (\ref{eq:LW}), and the differential scattering cross-section $\sigma'$ is defined from (\ref{eq:scat}) as 
\begin{equation}\label{eq:sigmap}
\sigma'(|\bk|,\theta')=\frac{8\pi|\bk|^4}{h}\sin^2(\theta'/2)\hat{R}\left(2|\sin(\theta'/2)\bk|\right),
\end{equation}
independent of the direction of $\bk$: the scattering is rotationally invariant because the flow is  isotropic.
Similarly, the total scattering cross-section (\ref{eq:tot_scat}) becomes
\begin{equation}\label{eq:Sigma_polar}
\Sigma(|\bk|)=\int_{-\pi}^{\pi}\sigma'(|\bk|,\theta)\, \d\theta.
\end{equation}

We now restrict our attention to horizontally homogeneous Wigner transforms, that is, NIWs whose large-scale properties are homogeneous. The group velocity term in (\ref{eq:transport}) vanishes and the equation for $W$ reduces to
\begin{equation}\label{eq:transport3}
\partial_t W=\mathcal{L}W-\Sigma(|\bk|)W.
\end{equation}
The solution of (\ref{eq:transport3}) can be calculated explicitly from the knowledge of the initial condition
\begin{equation}\label{eq:ICs_W_general}
W(|\bk|,\theta,t=0)=W_0(|\bk|,\theta)
\end{equation}
and the eigenvalues and eigenvectors of the operator $\mathcal{L}$. 
To find these, we remark that, from the symmetry property $\sigma'(|\bk|,\theta)=\sigma'(|\bk|,-\theta)$, (\ref{eq:L_theta}) can be viewed as a convolution of $\sigma'$ and $W$. Because the Fourier transform of a convolution is proportional to the product of the Fourier transforms of the convolved functions, the eigenvalues and eigenvectors of $\mathcal{L}$ are
\begin{equation}\label{eq:eigen}
\{\lambda_n,\cos(n\theta)\}, \ n=0,1,\cdots,
\end{equation}
where
\begin{equation}\label{eq:lambda_n}
\lambda_n=\int_{-\pi}^{\pi}\sigma'(|\bk|,\theta)\cos(n\theta) \, \d\theta
\end{equation}
is proportional to the Fourier transform of $\sigma'$ with respect to $\theta$ and depends on $|\bk|$. 
Since $\sigma'$ is non-negative and smooth,
\begin{equation}\label{eq:eigen1}
\lambda_0=\Sigma(|\bk|)\text{\hspace{.5cm} and \hspace{.5cm}}|\lambda_{n\geq 1}|<\lambda_0.
\end{equation}


Expanding the initial condition (\ref{eq:ICs_W_general}) in the basis of eigenfunctions given in (\ref{eq:eigen}), we write
\begin{equation*}
W_0(|\bk|,\theta)=\sum_{n=0}^{\infty}w_n(|\bk|)\cos(n\theta),
\end{equation*}
and obtain the exact solution of (\ref{eq:transport3}) as
\begin{equation}\label{eq:solution}
W(|\bk|,\theta,t)=\sum_{n=0}^{\infty}w_n(|\bk|)\e^{(\lambda_n-\Sigma)t}\cos(n\theta).
\end{equation}

From  (\ref{eq:eigen1}) and (\ref{eq:solution}), we conclude that, regardless of the initial conditions, the solution of (\ref{eq:transport3}) converges, at a given wavenumber $|\bk|$, to the stationary, isotropic solution given by 
\begin{equation}
W(|\bk|,\theta,t=\infty)=w_0(|\bk|)=\frac{1}{2\pi}\int_{-\pi}^{\pi}W(|\bk|,\theta,t=0) \, \d\theta.
\end{equation}
This is a key conclusion drawn from the transport equation: the scattering of NIWs by a random flow leads to an isotropic wave field. Moreover, the convergence timescale is deduced from (\ref{eq:solution}) to be approximately 
\begin{equation}\label{eq:isotropisation_time}
T\simeq 1/(\Sigma-\lambda'), \ \ \textrm{where}\ \ \lambda'=\max_{n \ge 1}\lambda_n.
\end{equation}
We note that the scattering time $\Sigma^{-1}$, which estimates the time scale over which scattering is significant, does not necessarily provide a reliable order of magnitude for the time scale of isotropisation (\ref{eq:isotropisation_time}) which can be much longer. This is demonstrated in section \ref{sec:gaussian} below.


\subsubsection{Solution for an intially plane wave}\label{sec:exact_solution}

To illustrate the isotropisation process, we consider an initial NIW-field that consists of a single plane wave, that is,  
\begin{equation}\label{eq:ICs_M}
M(\bx,t=0)=M_0 \e^{\i\bk_0 \cdot\bx},
\end{equation}
where $M_0$ is a constant.
Without loss of generality, we take $\bk_0$ to be aligned with the $x$-axis, $\mathbf{k_0}=(|\mathbf{k_0}|,0)$.
From (\ref{eq:Wig}), the Wigner transform corresponding to (\ref{eq:ICs_M}) is
\begin{equation}\label{eq:ICs_W}
W_0(\bk)=W(\bk,t=0)=\delta(\bk-\bk_0),
\end{equation}
or, equivalently, in polar coordinates,
\begin{equation}\label{eq:ICs_W1}
W_0(|\bk|,\theta)=|\bk_0|^{-1} \delta(|\bk|-|\bk_0|)\times\delta(\theta).
\end{equation}
The Wigner transform at later times can be similarly written in separable form,
\begin{equation}\label{eq:separable}
W(|\bk|,\theta,t)=|\bk_0|^{-1} \delta(|\bk|-|\bk_0|)W_\theta(\theta,t),
\end{equation}
where $W_\theta$ is solution of (\ref{eq:transport3}) with initial condition $W_\theta(\theta,t=0)=\delta(\theta)$. 
Using (\ref{eq:solution}) and the Fourier series for the Dirac function 
$$\delta(\theta)=\frac{1}{2\pi}+\frac{1}{\pi}\sum_{n=1}^\infty\cos(n\theta),$$ 
this can be written explicitly as 
\begin{equation}\label{eq:solution_single_harmonic}
W_\theta(\theta,t)=\frac{1}{2\pi}+\frac{1}{\pi}\sum_{n=1}^{\infty}\e^{(\lambda_n-\Sigma)t}\cos(n\theta).
\end{equation}
The influence of the background flow shows through the eigenvalues $\lambda_n$ and $\Sigma$, which are both functions of $|\bk|$. We compute these asymptotically and numerically for Gaussian flows in the next section.


\subsubsection{Gaussian random flow}\label{sec:gaussian}

We now take the streamfunction to be an isotropic, homogeneous Gaussian process characterised by its covariance $R(\bx)\propto\e^{-\kc^2|\bx|^2/2}$. The corresponding power spectrum is 
\begin{equation}
\hat{R}(|\bk|)=A \,\e^{-|\bk|^2/(2\kc^2)},
\label{eq:power}
\end{equation} for some $A>0$, and implies a correlation length $\lc=2\sqrt{2\pi}/\kc$ with the definition we have chosen \footnote{The correlation length $\lc$ is defined as $\lc=2\pi/K_\mathrm{c}$, where $K_\mathrm{c}=\iint |\bk|\hat{R}(\bk) \, \d \bk /\iint\hat{R}(\bk) \, \d \bk$ is the correlation wavenumber.}. 
For this spectrum, it is instructive to calculate approximations to the eigenvalues (\ref{eq:lambda_n}) and hence deduce the isotropisation time scale (\ref{eq:isotropisation_time})  in the two limiting cases $|\bk|\ll \kc$ and $|\bk|\gg \kc$. 

First, for  $|\bk|\ll \kc$, $\hat{R}\sim 1$ in (\ref{eq:sigmap}) and $\sigma'(\theta')\propto\sin^2(\theta'/2)$; in this regime, corresponding to waves of scales much larger than the flow correlation scale,
\begin{equation}\label{eq:lambda_01}
\lambda_0=\Sigma \sim \frac{8\pi^2|\bk|^4A}{h},
\end{equation}
while 
$\lambda_1 \sim -\lambda_0/2$ and $\lambda_{n\geq2}=o(\lambda_0)$.
Therefore, the isotropisation time-scale is approximately $\Sigma^{-1}$.

\begin{figure}
\begin{center}
\includegraphics[width=25pc,angle=0]{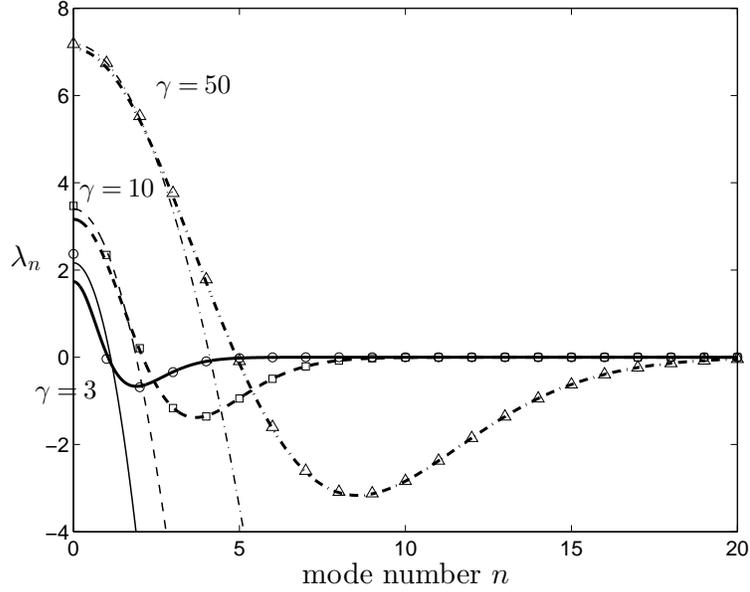}
\caption{Eigenvalues of the scattering cross-section operator $\mathcal{L}$ obtained numerically (symbols) and asymptotically from (\ref{eq:lambda_n_formula}) (thick lines) and (\ref{eq:lambda_n_formula_1}) (thin lines) for $\gamma=3$ ($\circ$ and solid line), $10$ ($\Box$ and dashed line) and $50$ ($\triangle$ and dashed-dotted line).
The eigenvalues are scaled by the factor $2\pi^{3/2}\kc^4A/h$ appearing in (\ref{eq:lambda_n_formula}) and (\ref{eq:lambda_n_formula_1}).}
\label{fig:lambda_n}
\end{center}
\end{figure}

Second, for $|\bk|\gg \kc$, using a steepest-descent method (detailed in Appendix \ref{app:steepest1}), we find
\begin{equation}\label{eq:lambda_n_formula}
\begin{split}
\lambda_n\sim & \frac{2\pi^{3/2}\kc^4A}{h}\gamma^{3/2}\exp\left(\gamma(\alpha_n^{1/2}-1)/{2}-n\sinh^{-1}(2n/\gamma)\right) \\
&\times \left( (\alpha_n^{-1/4}-\alpha_n^{1/4})+\frac{1}{6\gamma}
(\alpha_n^{-1/ 4}-\alpha_n^{-3/4}+\frac{7}{2}\alpha_n^{-5/4}+\frac{5}{2}\alpha_n^{-7/4})
\right),
\end{split}
\end{equation} 
where $\gamma=2|\bk|^2/\kc^2\gg 1$, $\alpha_n=1+4n^2/\gamma^2$, and we have assumed that $n=O(\gamma)$ to obtain an approximation uniformly valid for large $n$. 
Expression (\ref{eq:lambda_n_formula}) is plotted in Fig.\ \ref{fig:lambda_n} as a thick line for three values of $\gamma$, along with the exact value (\ref{eq:lambda_n}) calculated numerically. The approximation (\ref{eq:lambda_n_formula}) is accurate for $n \ge 1$, even for the moderately large $\gamma=3$, but it fails for $n=0$ because of the assumption $n=O(\gamma)$ breaks down. An application of Laplace's method for $n=O(1)$ detailed in Appendix \ref{app:steepest2}, gives the approximation
\begin{equation}\label{eq:lambda_n_formula_1}
\lambda_n\sim \frac{2\pi^{3/2}\kc^4A}{h}\gamma^{1/2}\left(1+\gamma^{-1}({3}/{4}-3n^2)\right).
\end{equation}
in place of (\ref{eq:lambda_n_formula}). Expression (\ref{eq:lambda_n_formula_1}) is plotted in Fig.\ \ref{fig:lambda_n} as a thin line for the three values of $\gamma$ used above.
The leading order of (\ref{eq:lambda_n_formula_1}) agrees with (\ref{eq:lambda_n_formula}) for $n=0$. Moreover, from (\ref{eq:lambda_n_formula_1}), $\lambda_n$ decreases with $n$ for $n=O(1)$. Combining the results (\ref{eq:lambda_n_formula_1}) and (\ref{eq:lambda_n_formula}) holding for different ranges of $n$, we deduce that $\lambda_0$ and $\lambda_1$ are the two largest eigenvalues of $\mathcal{L}$, with $\lambda_1<\lambda_0$. Consequently, for $|\bk|\gg \kc$, the scattering and isotropisation time-scales are, from (\ref{eq:lambda_n_formula_1}), 
\begin{equation}\label{eq:scat_time_adv}
\Sigma^{-1}\sim\frac{h}{2\pi^{3/2}\kc^4A}\gamma^{-1/2}.
\end{equation}
and
\begin{equation}\label{eq:isotrop_adv}
(\Sigma-\lambda_1)^{-1}\sim\frac{h}{6\pi^{3/2}\kc^4A}\gamma^{1/2}.
\end{equation}
Therefore, as $\gamma\rightarrow\infty$, that is, as the wave scales decrease, the scattering time-scale decreases as $\gamma^{-1/2}$ while the isotropisation time-scale increases as $\gamma^{1/2}$. Thus NIWs with scales much smaller than the flow scales are quickly impacted by the flow but require long times to  isotropise fully.


\subsection{Importance of advection}

It is interesting to examine the specific effect of advection on the scattering as this process is absent from the Schr\"odinger equation treated in \citep{Ryzhik96}. We do this by neglecting advection and analysing how the various quantities calculated in the previous section change.
Without advection, the term proportional to $|\bk\times\bp|^2$ disappears from (\ref{eq:scat}) (see  (\ref{eq:transport1})--(\ref{eq:defV})), and (\ref{eq:sigmap}) becomes
\begin{equation}\label{eq:sigmap1}
\sigma'(|\bk|,\theta')=\frac{8\pi|\bk|^4}{h}\sin^4({\theta'}/{2})\hat{R}(2|\bk|\sin(\theta'/2)).
\end{equation}
Let us now see how the change of exponent in (\ref{eq:sigmap1}) impacts on the isotropisation timescale in the two regimes studied above. In the case $|\bk|\ll \kc$, it is easy to show that
$$\Sigma=\frac{6\pi^2|\bk|^4A}{h},$$
while 
$\lambda_1=-2\lambda_0/3$, $\lambda_2=\lambda_0/6$ and $\lambda_{n\geq3}=o(\lambda_0)$.
Therefore, the isotropisation timescale is
\begin{equation}\label{eq:lambda_02}
(\Sigma-\lambda_2)^{-1}=\frac{h}{5\pi^2|\bk|^4A}.
\end{equation}
In this case, advection only decreases the isotropisation timescale by the modest factor $5/8$ (compare the inverse of (\ref{eq:lambda_01}) with (\ref{eq:lambda_02})).
Its impact is more dramatic in the case $|\bk|\gg \kc$. Without advection, it is possible to show, using similar techniques as those of Appendix \ref{app:steepest}, that the two largest eigenvalues are $\lambda_0$ and $\lambda_1$, so that the scattering and isotropisation timescales are
\begin{equation}\label{eq:scat_time_no_adv}
\Sigma^{-1}\simeq\frac{h}{3\pi^{3/2}\kc^4A}\gamma^{1/2}
\end{equation}
and
\begin{equation}\label{eq:isotrop_no_adv}
(\Sigma-\lambda_1)^{-1}=\frac{h}{15\pi^{3/2}\kc^4A}\gamma^{3/2}.
\end{equation}
Thus advection dramatically decreases the scattering and isotropisation timescales, by factors proportional to $1/\gamma\propto1/|\bk|^2$ (compare (\ref{eq:scat_time_adv})--(\ref{eq:isotrop_adv}) and (\ref{eq:scat_time_no_adv})--(\ref{eq:isotrop_no_adv})). 
This is confirmed by numerical simulations (not shown).

These results can be understood physically by considering the YBJ equation (\ref{eq:YBJ}). The regime $|\bk|\ll \kc$ corresponds to $\lw/\lf\gg1$, when advection is much smaller than refraction and has little influence the dynamics.
In contrast, for $|\bk|\gg \kc$ or $\lw/\lf \ll1$, advection dominates over refraction and its effect considerably accelerates the scattering process.


\subsection{Numerical simulations}
In this section, we  validate the theoretical findings of the previous section using numerical simulations of the YBJ equation (\ref{eq:YBJ}) for an initially plane wave. We focus particularly on the isotropisation time scale.

\subsubsection{Quantifying the isotropisation}
We consider the evolution of a wavefield that is initially strongly anisotropic, with initial condition (\ref{eq:ICs_M}), corresponding to (\ref{eq:ICs_W}) or (\ref{eq:ICs_W1}) in terms of  Wigner transform. To analyse  its scattering, we calculate the Fourier transform of $M$ rather than its Wigner transform. 
This has two advantages. 
First, from the dimensional version of (\ref{eq:Wig_dual}),
\begin{equation}\label{eq:Wig_dual_dim}
W(\bx,\bk,t)=\int_{\mathbb{R}^2} \e^{\i\bp\cdot\bx}\hat{M}(-\bk-\bp/2,t)\hat{M}^*(-\bk+{\bp}/{2},t) \, \d\bp,
\end{equation}
where $\hat{M}$ denotes the  Fourier transform of $M$, it follows that 
\begin{equation}\label{eq:linkMW}
|\hat{M}(\bk,t)|^2=4\pi^2\int_{\mathbb{R}^2} W(\bx,-\bk,t) \, \d\bx.
\end{equation}
(The minus sign arises from our particular definition of the Fourier transform, see (\ref{eq:Fourier_def}).)
According to the perturbation theory of Appendix \ref{app:deriv}, the leading-order Wigner transform is spatially independent and the higher-order terms have vanishing spatial averages  when there is no large-scale dependence on the initial condition and velocity field, as assumed here (see (\ref{eq:expansion_W})). Therefore, 
$|\hat{M}(\bk,t)|^2$ directly measures the leading-order Wigner transform. 
Second, the analysis is obviously much easier with the reduction from the five dimensions of the Wigner transform to the three dimensions of the Fourier transform.

A natural measure of the isotropy of the wave field is the ratio $r$ of the kinetic energy associated with the right-hand part of the Fourier spectrum ($k>0$) to the total kinetic energy, 
\begin{equation}\label{eq:ratio_isotrop}
r(t)=\frac{1}{E}\iint_{k>0,l}|\hat{M}(\bk,t)|^2 \, \d \bk,
\end{equation}
where $E$ is the integral appearing in (\ref{eq:ratio_isotrop}) extended over the entire $(k,l)$-plane. An isotropic field is thus characterized by $r=0.5$ and the initial condition (\ref{eq:ICs_M}) by $r(0)=0$.
We choose this measure over more sophisticated ones because (i) it is easy to calculate without the need to resort to polar coordinates, (ii) the large domain of integration (a half plane in $(k,l)$) gives a  measure that is smooth in time.

As described in section \ref{sec:exact_solution}, the initial condition (\ref{eq:ICs_W}) leads to an exact solution given by (\ref{eq:solution_single_harmonic})--(\ref{eq:separable}) for the Wigner transform. Using this and (\ref{eq:linkMW}), we compute the ratio $r$ for this solution as
\begin{align}
r(t)
&=\frac{4\pi^2 S}{E}\int_{\pi/2}^{3\pi/2}W_\theta(\theta,t) \, \d\theta  \nonumber \\
&=\frac{1}{2}-\frac{2}{\pi}\sum_{n=0}^\infty\frac{(-1)^n}{2n+1} \e^{-(\Sigma-\lambda_{2n+1})t}, \label{eq:exact_r}
\end{align}
where $S$ is the area of the domain. The second term in (\ref{eq:exact_r}) tends to $0$ as $t\rightarrow\infty$.

\begin{figure}
\begin{center}
\includegraphics[width=.35\textwidth]{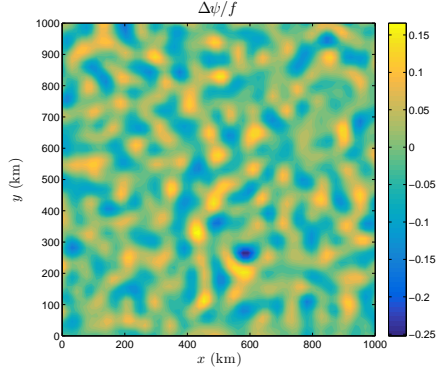}
\caption{Vorticity field, normalised by the Coriolis frequency $f$, for one  realisation of the Gaussian random streamfunction with power spectrum (\ref{eq:power}) and correlation length $200$ km.}
\label{fig:vort_f}
\end{center}
\end{figure}

\subsubsection{Parameters and results}
Equation (\ref{eq:YBJ}) is solved on a doubly periodic $512\times512$ grid using a
pseudo-spectral time-split Euler scheme. A weak biharmonic
dissipation is added for numerical stability. The streamfunction
$\psi$ is taken as a realization of a homogeneous isotropic 
Gaussian random field, with  power spectrum (\ref{eq:power}) with the correlation length $\lc=200$ km. It is time-independent but, according to Appendix \ref{app:deriv}, any time dependence with stationary statistics would lead to the same results provided that it is slow compared with the wave time scale. 
The domain is a square of length $40\lc=8000$ km. The Coriolis frequency is $f_0=10^{-4}$s$^{-1}$ and the Rossby radius of deformation is $r_d=20$ km, representative of the North Atlantic; this gives a dispersion parameter $h=f_0r_d^2=4\cdot10^{4} \, \text{m}^2 \, \text{s}^{-1}$. We verify that, because $(f_0^2+f_0h(2\pi/\lc)^2)^{1/2}-f_0\simeq 2\cdot10^{-5}\ll f_0$, waves with a similar wavelength as the background flow satisfy the near-inertial approximation. 
The amplitude $A$ of the power spectrum is chosen such that 
the root-mean-square of the vorticity field is  $\zeta_{\text{rms}}\simeq 5\cdot10^{-6}$s$^{-1}\ll f_0$. Thus, the parameter $\Psi/h=(\lc/(2\pi))^2\zeta_{\text{rms}}/h\simeq 0.12$ is much smaller than $1$, in accordance with the scaling used in Appendix \ref{app:deriv}.
A typical subdomain of size $1000 \, \text{km}\times 1000 \, \text{km}$ of the vorticity field is shown in Fig.\ \ref{fig:vort_f}.

\begin{figure}
\begin{center}
\begin{tabular}{cc}
(a)\raisebox{-.95\height}{\includegraphics[width=.35\textwidth]{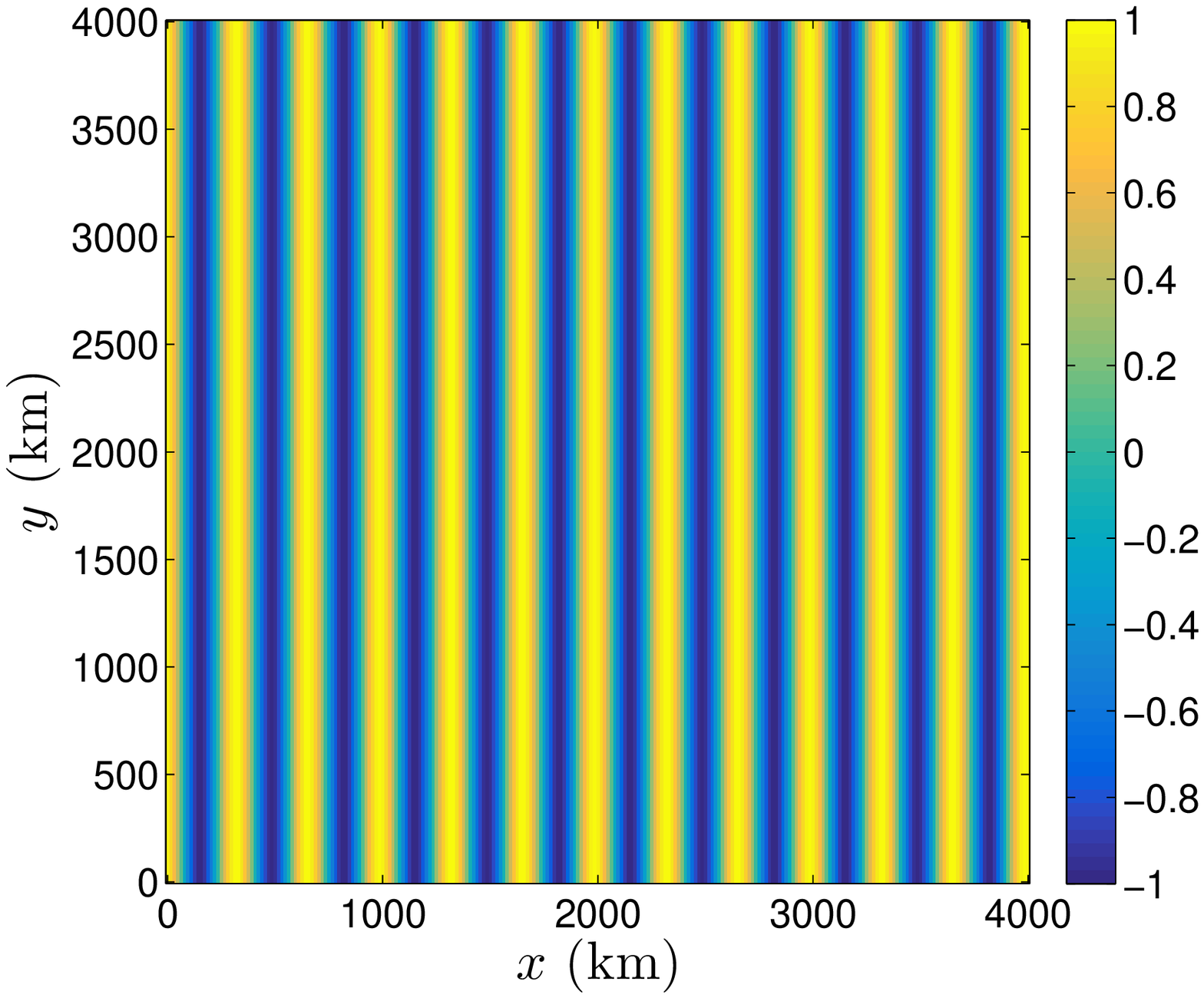}}&
(b)\raisebox{-.95\height}{\includegraphics[width=.35\textwidth]{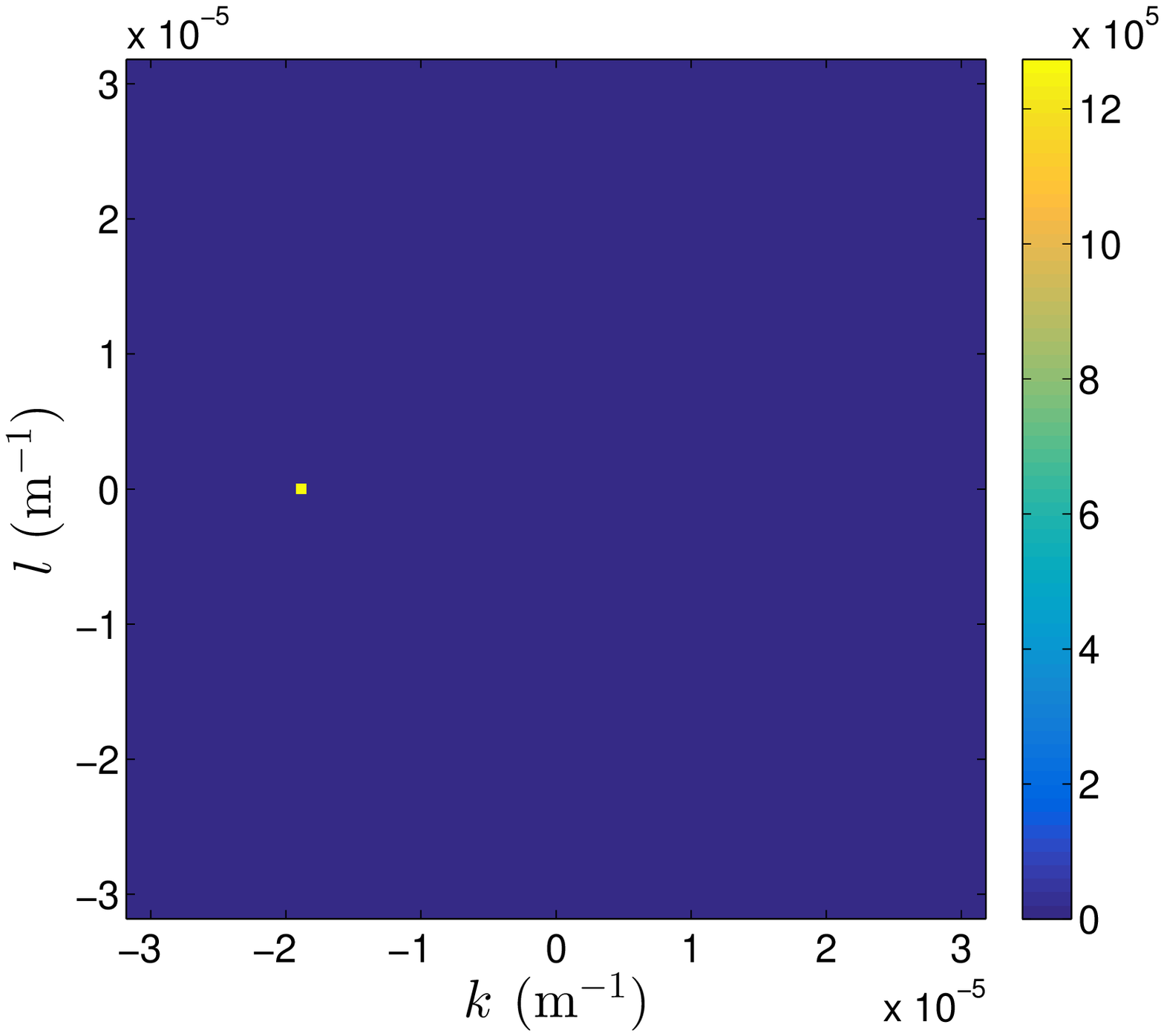}}\\
(c)\raisebox{-.95\height}{\includegraphics[width=.35\textwidth]{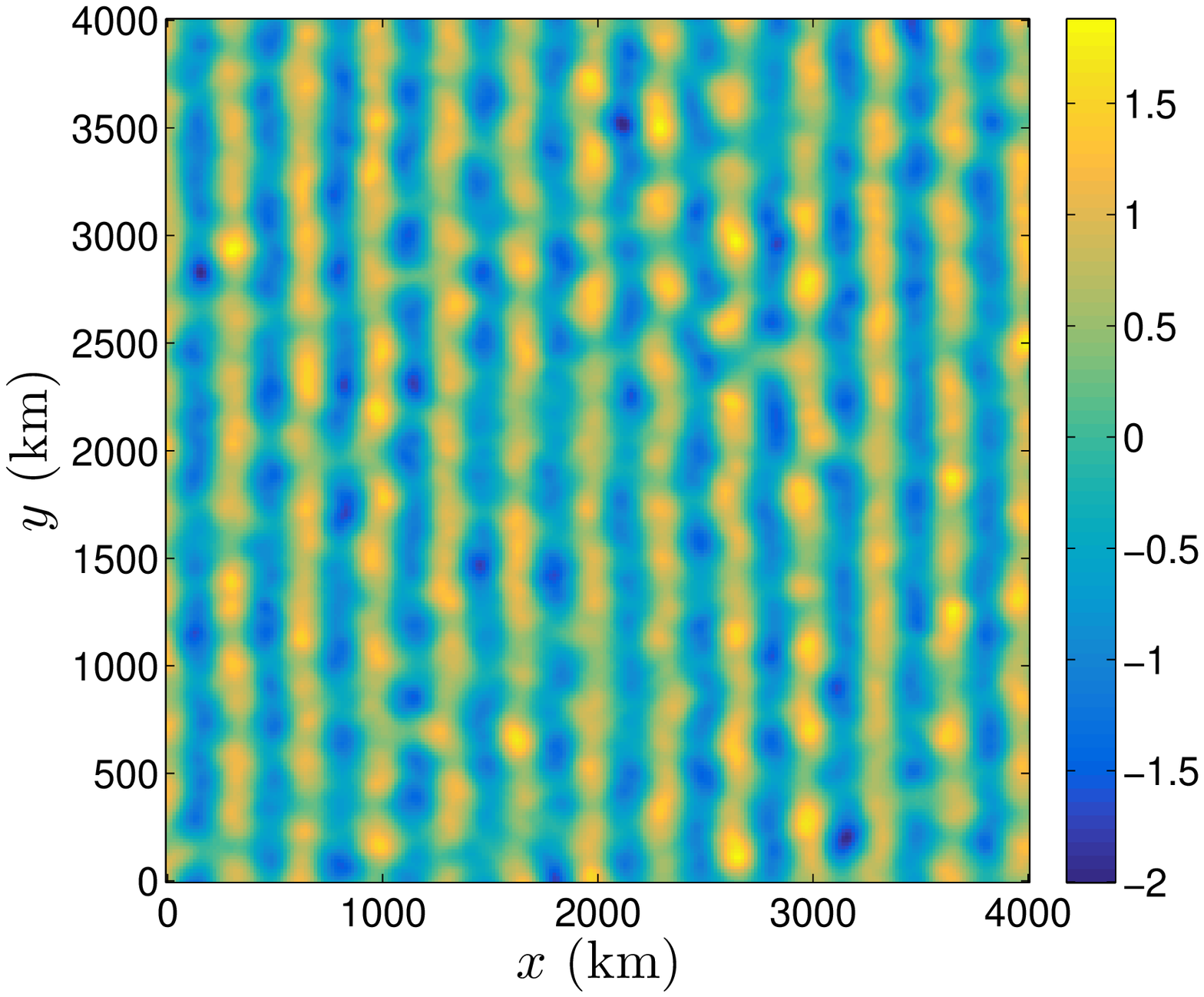}}&
(d)\raisebox{-.95\height}{\includegraphics[width=.35\textwidth]{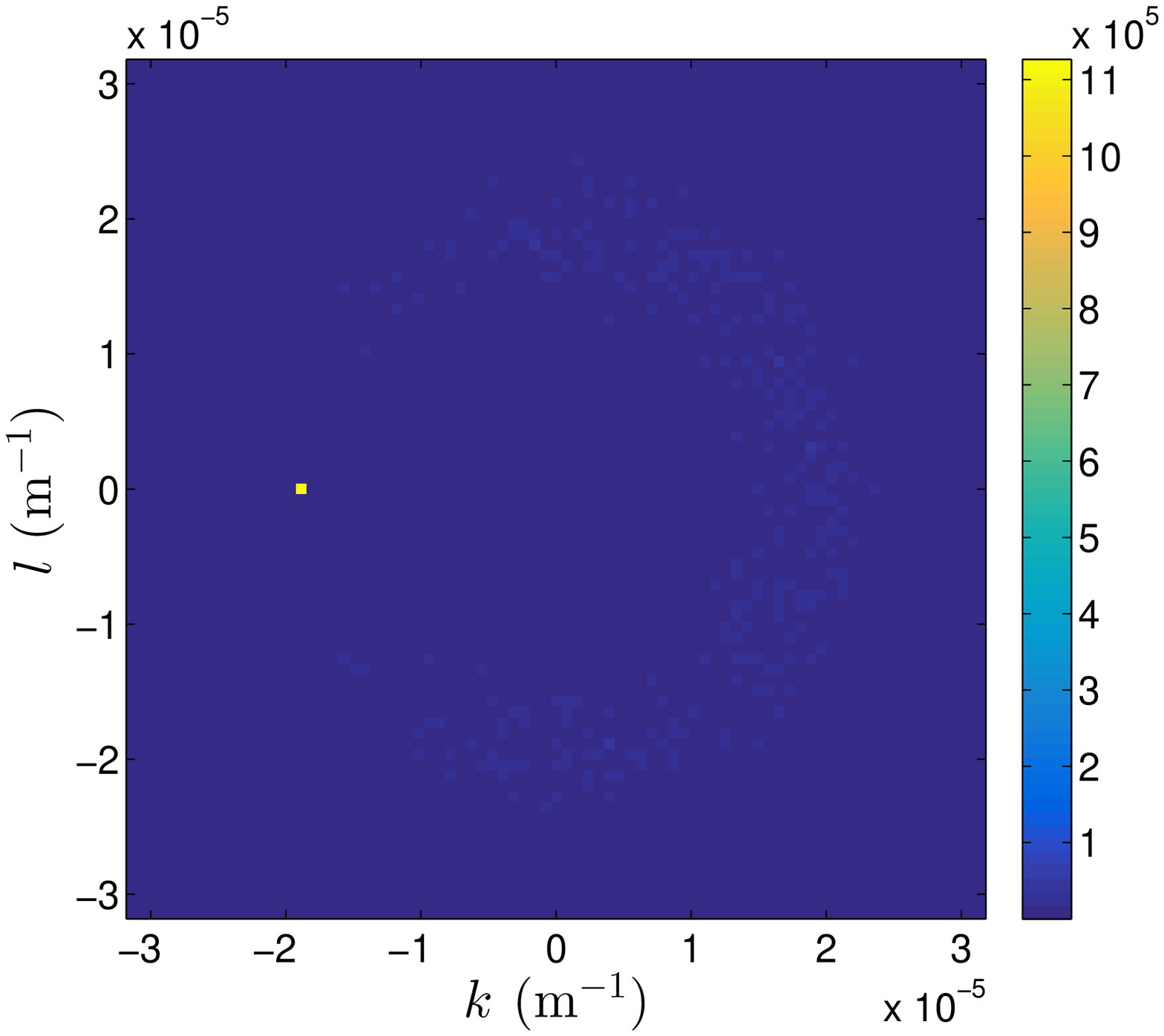}}\\
(e)\raisebox{-.95\height}{\includegraphics[width=.35\textwidth]{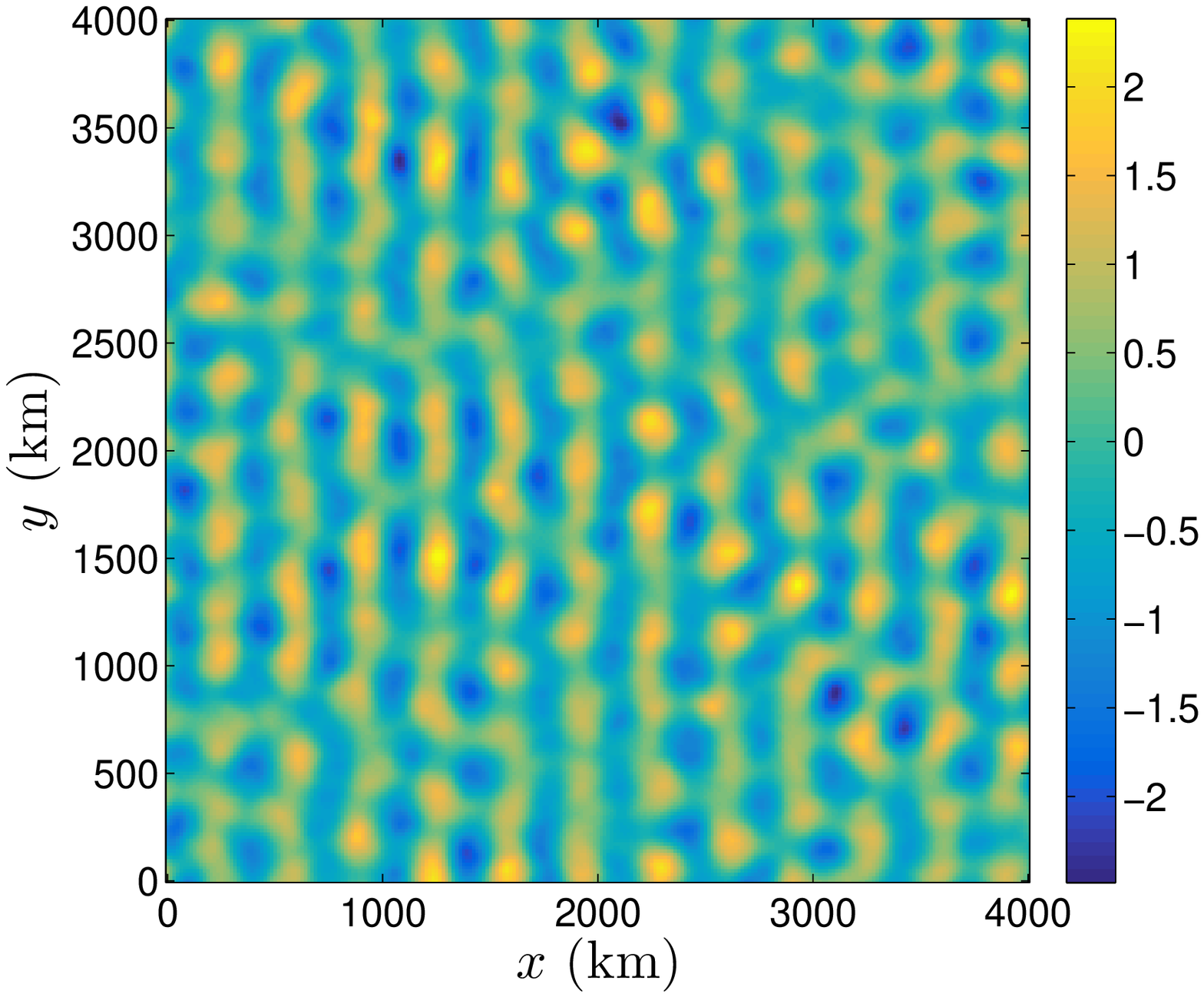}}&
(f)\raisebox{-.95\height}{\includegraphics[width=.35\textwidth]{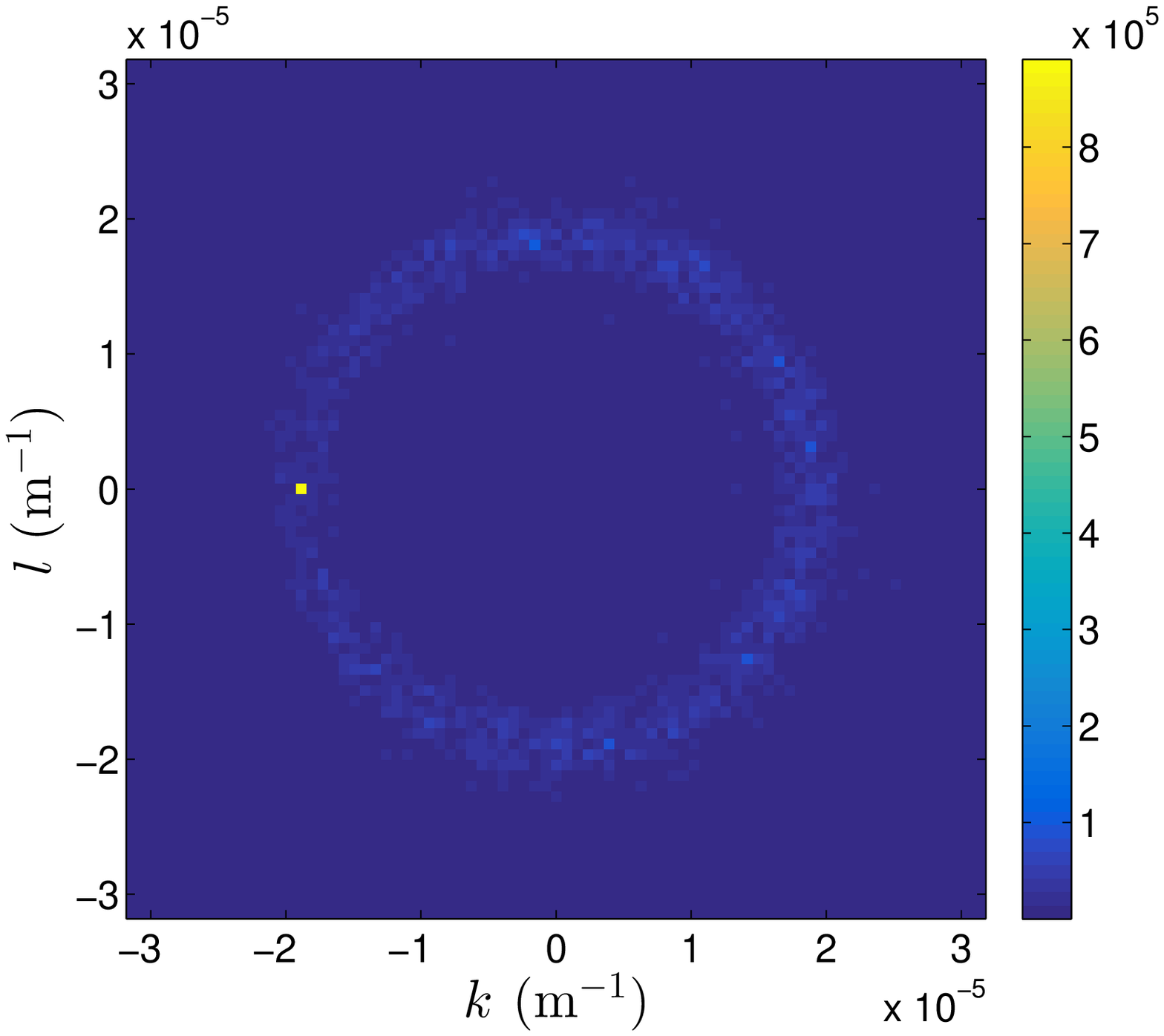}}\\
(g)\raisebox{-.95\height}{\includegraphics[width=.35\textwidth]{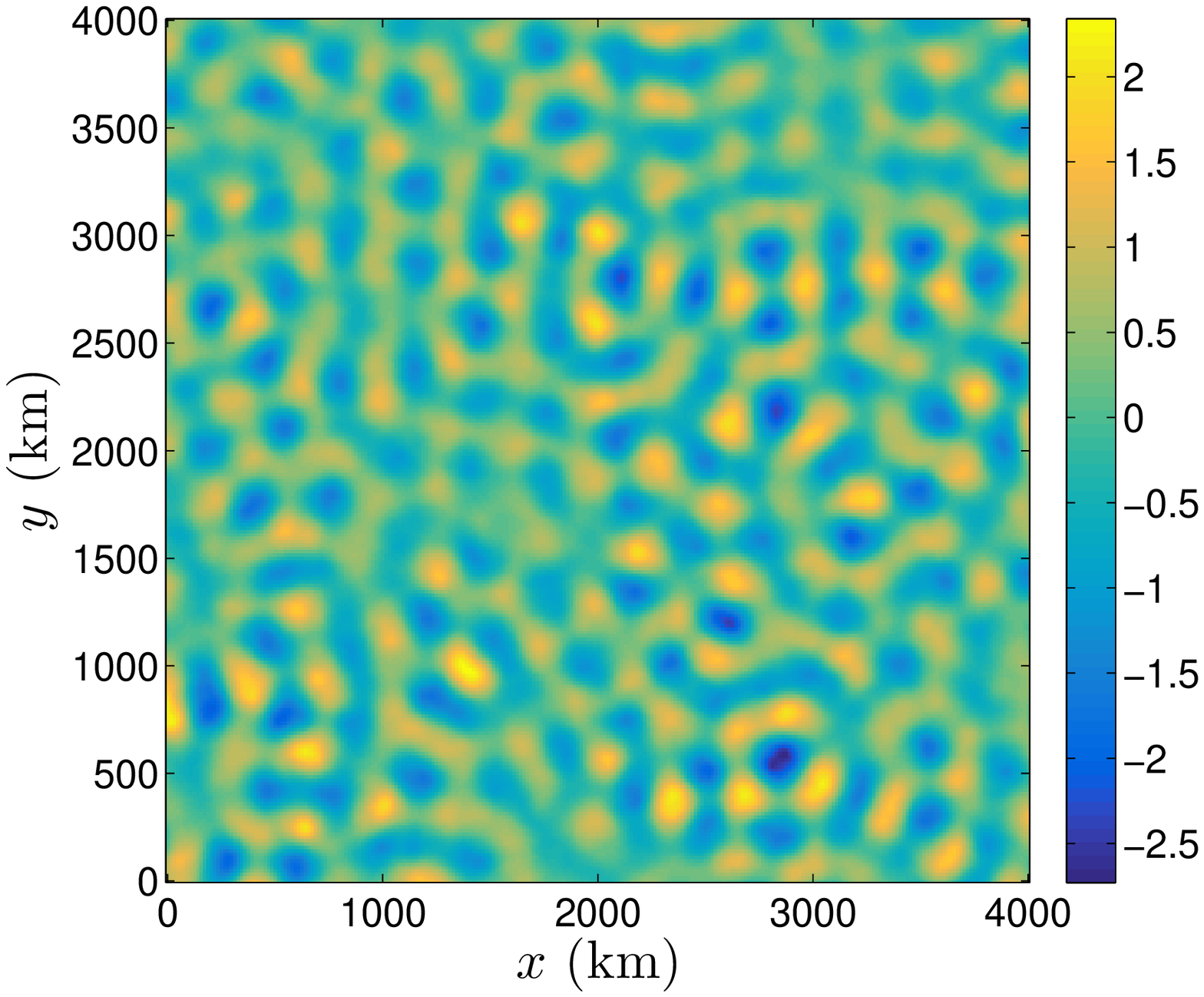}}&
(h)\raisebox{-.95\height}{\includegraphics[width=.35\textwidth]{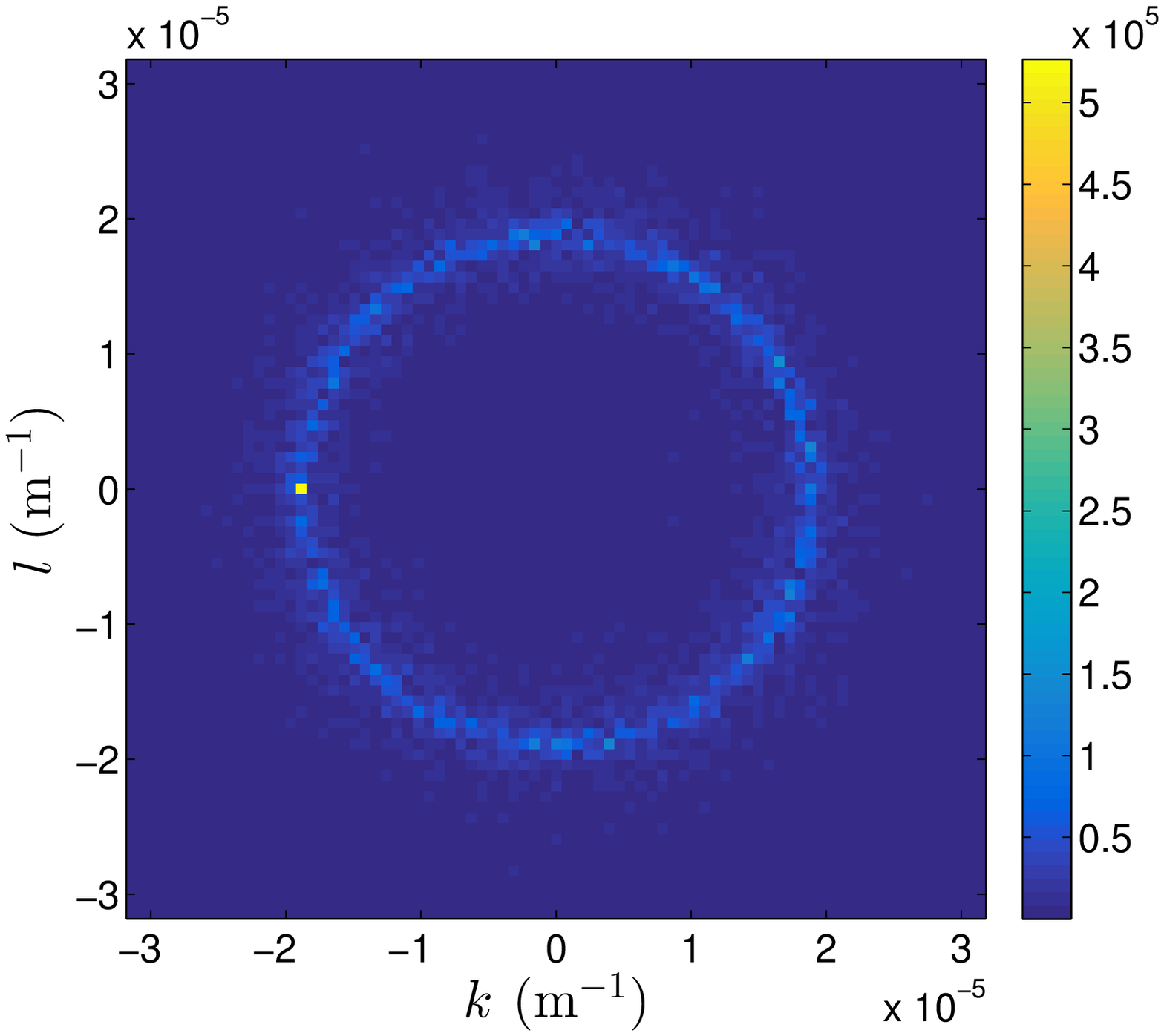}}
\end{tabular}
\caption{Evolution of  $\Re M $ in a $4000$ km$\times4000$ km subdomain (left), and of the Fourier transform amplitude $|\hat{M}|$ for an initially plane wave, shown at $t=0$ (a, b), $10$ days (c, d), $30$ days (e, f) and $75$ days (g, h) for $\gamma=1.1$ (see text for the value of other parameters).}
\label{fig:time_evolution}
\end{center}
\end{figure}

Three different values of the parameter $\gamma=2|\bk_0|^2/\kc^2$ are used: $\gamma=0.5$, $1.1$ and $4.9$, corresponding to a NIW wavelength $2\pi/|\bk_0|$ of approximately $500$, $340$ and $160$ km.
Fig.\ \ref{fig:time_evolution} displays the  evolution of $M$, specifically $\Re M$  alongside the magnitude  of its Fourier transform $|\hat{M}|$, for $\gamma=1.1$. As expected, the initially unidirectional wave field (Fig.\ \ref{fig:time_evolution}a) is slowly modulated by the flow, leading to the generation of fluctuations in all directions (Figs.\ \ref{fig:time_evolution}c, e), and ends up almost isotropic  (Fig.\ \ref{fig:time_evolution}g). This is confirmed by the amplitude of the Fourier transform $|\hat{M}|$ (Fig.\ \ref{fig:time_evolution}, right column), which starts from a single point $(-|\bk_0|,0)$ in the $(k,l)$-plane (Fig.\ \ref{fig:time_evolution}b), then develops into a thin annulus of radius $|\bk_0|$ as the energy at $(-|\bk_0|,0)$ decreases (Figs.\ \ref{fig:time_evolution}d, f, h).

\begin{figure}
\begin{center}
\begin{tabular}{ccc}
\includegraphics[width=.3\textwidth]{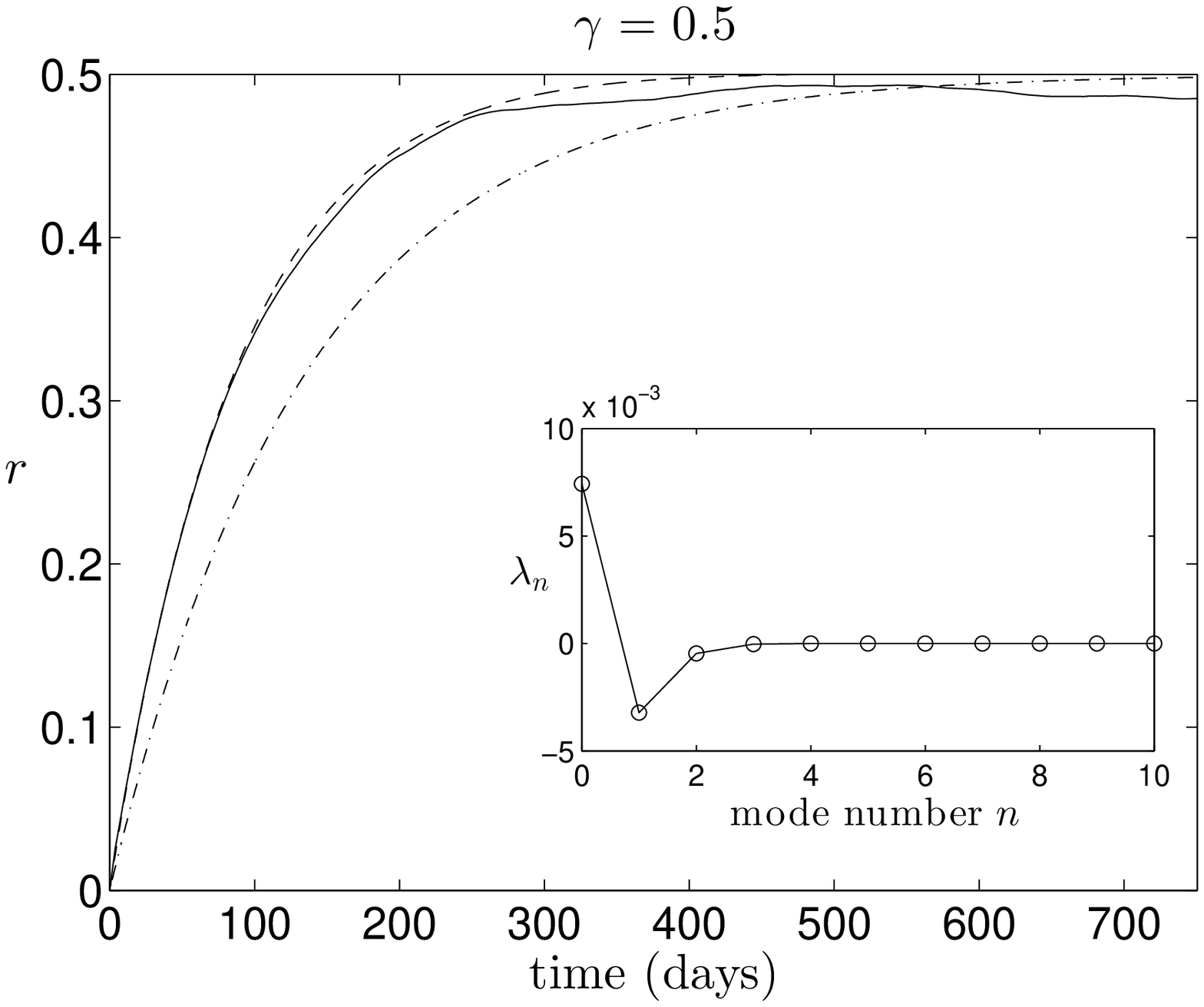}&
\includegraphics[width=.3\textwidth]{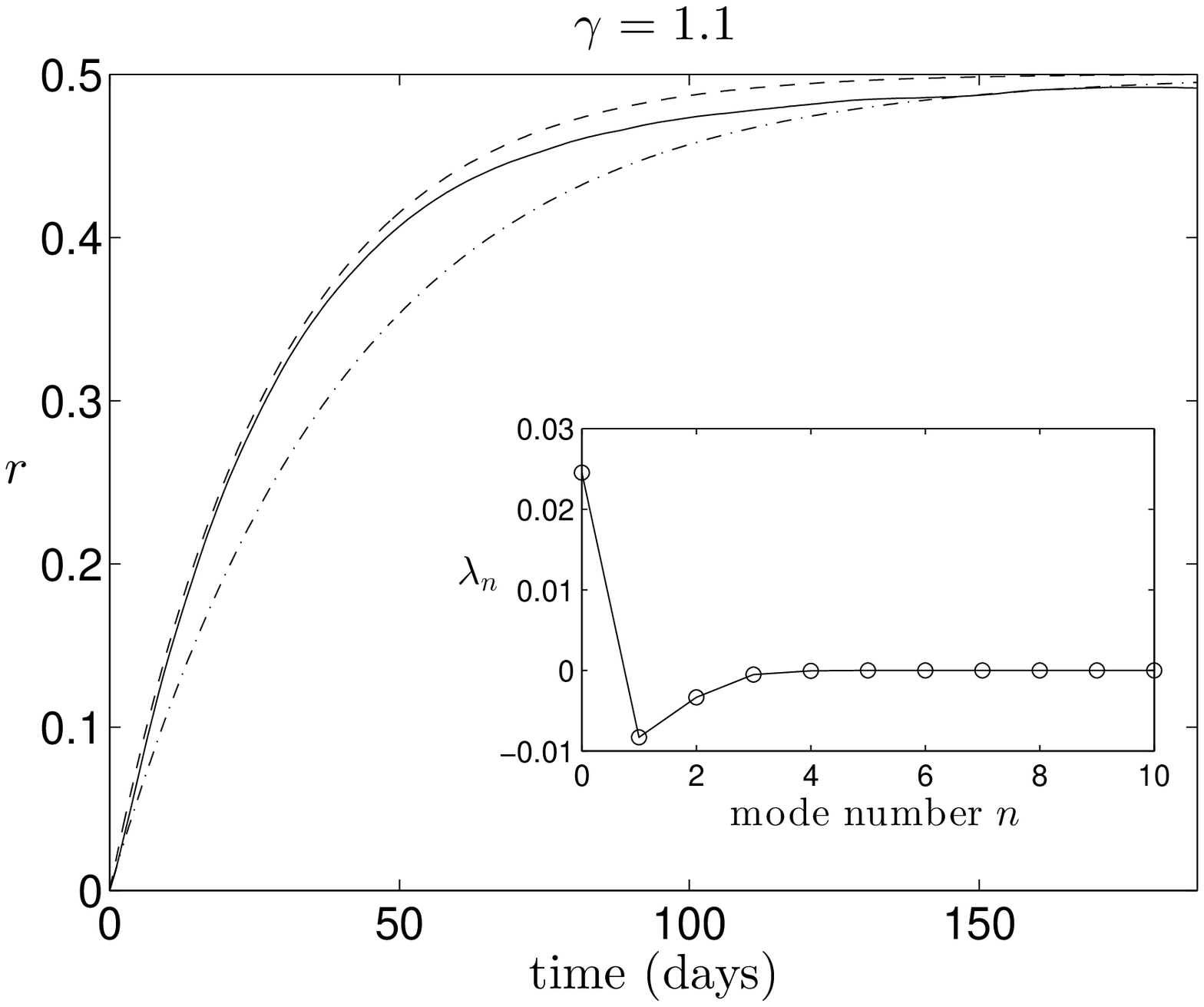}&
\includegraphics[width=.3\textwidth]{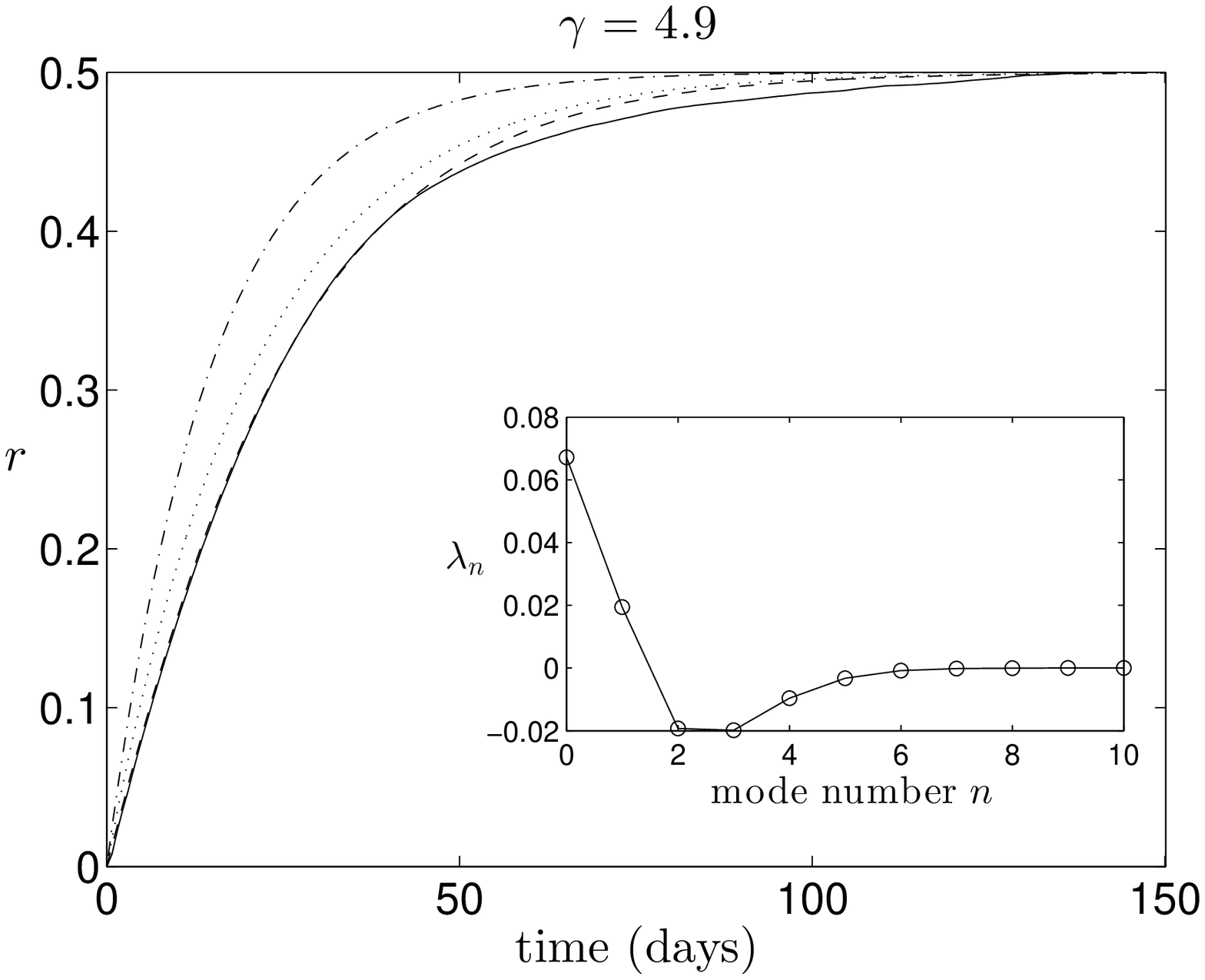}
\end{tabular}
\caption{Evolution of the energy ratio $r$ defined in (\ref{eq:ratio_isotrop}) for  $\gamma=2k_0^2/\kc^2= 0.5$ (left), 1.1 (middle) and 4.9 (right). The average ratio calculated from $20$ simulations for each value of $\gamma$ (solid lines) is compared with the  theoretical prediction (\ref{eq:exact_r}) (dashed lines) and with the approximations $r_\Sigma=(1-\exp(-\Sigma t))/2$ (dash-dotted lines) and $r_{\lambda'}=(1-\exp(-(\Sigma-\lambda')t))/2$ (dotted lines). For $\gamma=0.5$ and $1.1$, $r_{\lambda'}$ is indistinguishable from $r_\Sigma$. Note the different time ranges in each figure. The eigenvalues $\lambda_n$ defined in (\ref{eq:lambda_n}) are shown in the insets. Units of $r$ and $\lambda_n$ are days$^{-1}$.}
\label{fig:ratio}
\end{center}
\end{figure}

For each value of $\gamma$, simulations like the one used for Fig.\ \ref{fig:time_evolution} have been repeated $20$ times with different realisations of the streamfunction. From these simulations, we calculate the average  of the ratio $r(t)$ defined in (\ref{eq:ratio_isotrop}). Fig.\ \ref{fig:ratio} displays the  evolution of this ratio for the three values of $\gamma$, together with the theoretical prediction (\ref{eq:exact_r}) and the two estimates
\begin{equation}
r_\Sigma(t)=\frac{1}{2}\left(1-\e^{-\Sigma t}\right)\text{ and }r_{\lambda'}(t)=\frac{1}{2}\left(1-\e^{-(\Sigma-\lambda')t}\right),
\end{equation}
where $\lambda'$ is defined in (\ref{eq:isotropisation_time}). These are crude estimates based on the scattering and isotropisation time scales $\Sigma^{-1}$ and $(\Sigma-\lambda')^{-1}$. 
The agreement between the numerical simulations (solid line) and the theoretical prediction (dashed line) is excellent, considering the strong assumptions underlying the derivation of (\ref{eq:transport3}).
The estimates $r_\Sigma$ and $r_{\lambda'}$ offer good, albeit less accurate, approximations. In particular, for $\gamma=4.9$, $r_{\lambda'}$ offers a significant improvement over $r_\Sigma$, highlighting the fact that $(\Sigma-\lambda')^{-1}$ is a more appropriate time-scale for isotropisation than $\Sigma^{-1}$.

From Fig.\ \ref{fig:ratio}, the isotropisation e-folding time-scale is approximately $81$, $27$ and $23$ days for $\gamma=0.5$, $1.1$ and $4.9$.
This is in agreement with expression (\ref{eq:lambda_01}) for small $\gamma$: as $\gamma$ or $|\bk|$ decreases, the time-scale increases. For $\gamma$ larger than $4.9$, we expect the isotropisation time-scale to increase again, in agreement with (\ref{eq:isotrop_adv}). The relatively short isotropisation time scale  found for $\gamma=1.1$ and $4.9$ is comparable with other time scales relevant to NIW dynamics, such as the vertical propagation time scale (see Ref.\ \citenum{YBJ}) -- a phenomenon  absent here because of the simplified reduced-gravity shallow-water setup -- or the $\beta$-dispersion time scale $|\bk|/\beta$. 

\section{Applications}\label{sec:storm}
 
We now consider two physical scenarios in which scattering plays an important role.
Scattering has the most dramatic impact on  waves that would be unidirectional in the absence of flow, as it then leads to dispersion in multiple directions instead. Examples of such waves in the ocean are inertial waves propagating equatorward due to the $\beta$-effect, and inertial waves generated by a moving cyclone.
We examine these two types of waves in what follows, using simulations of a reduced gravity shallow-water model of the mixed-layer (cf.\ Ref.\ \citenum{Klein04}, section 4). This model reduces to the YBJ model of the previous sections when the near-inertial and linear approximations are made. By relaxing these approximations, we confirm the relevance of the theoretical results to the more realistic setup. 
We continue to assume a time-independent flow, an assumption that should not alter the results provided the flow evolves slowly enough. 

\subsection{Wavepacket on a $\beta$-plane}

For large spatial and time scales, the $\beta$-effect associated with the earth's curvature becomes important. As a result, waves develop a zonally banded structure characterized by negative meridional wavenumbers and  equatorward propagation, as described by a ray-tracing (WKB) approximation. Such waves are ubiquitous in global simulations \citep{Komori08,Blaker2012} and oceanic measurements \citep{Dasaro95}. Although this process is well understood in simple configurations \citep{Buhler2003}, the additional effect of a background flow remains, to our knowledge, unexplored. 
As we shall see, this effect can, under some circumstances, strongly affect the wave properties.

\begin{figure}
\begin{center}
\begin{tabular}{cc}
(a)\raisebox{-.9\height}{\includegraphics[width=.45\textwidth]{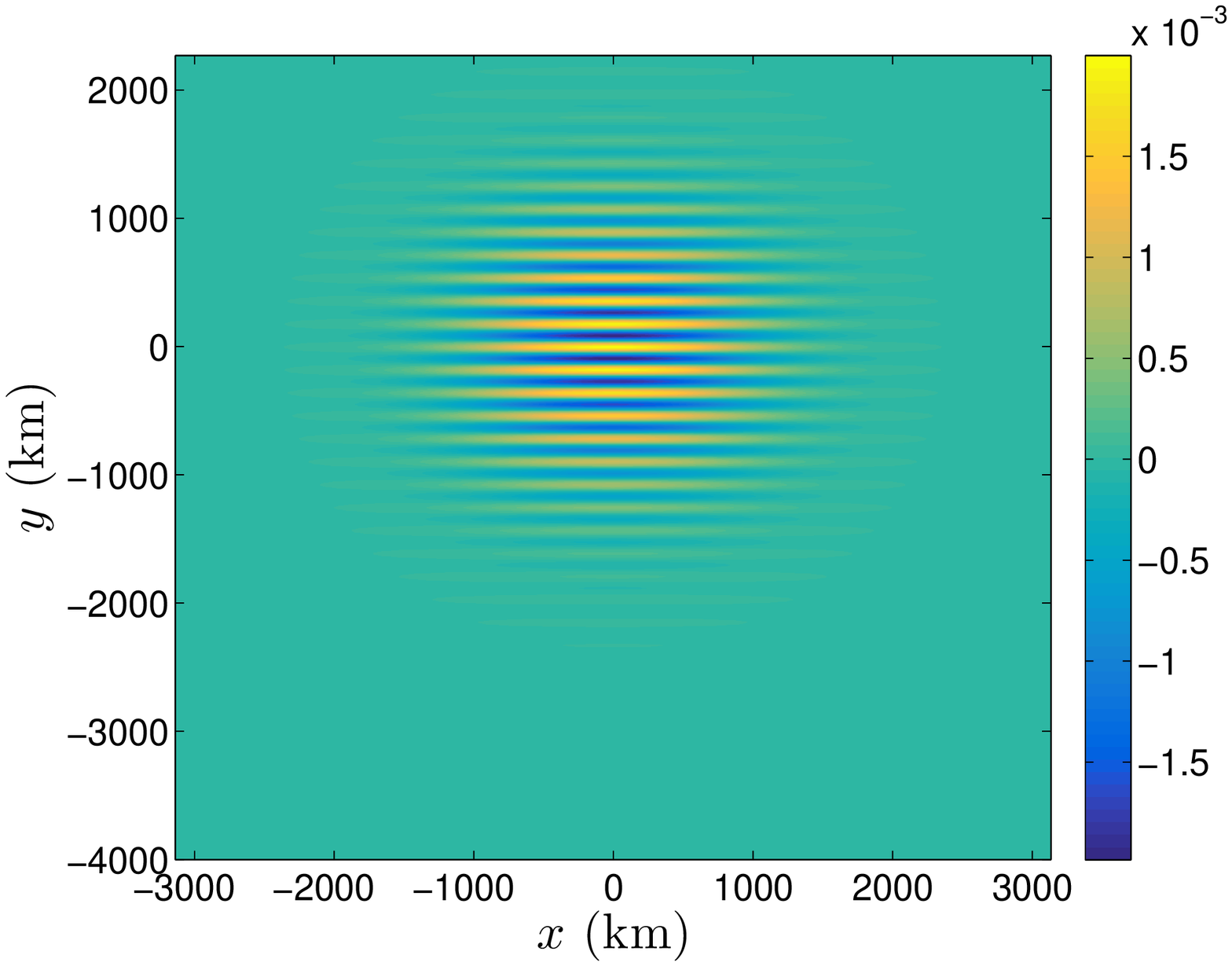}}&
(b)\raisebox{-.9\height}{\includegraphics[width=.45\textwidth]{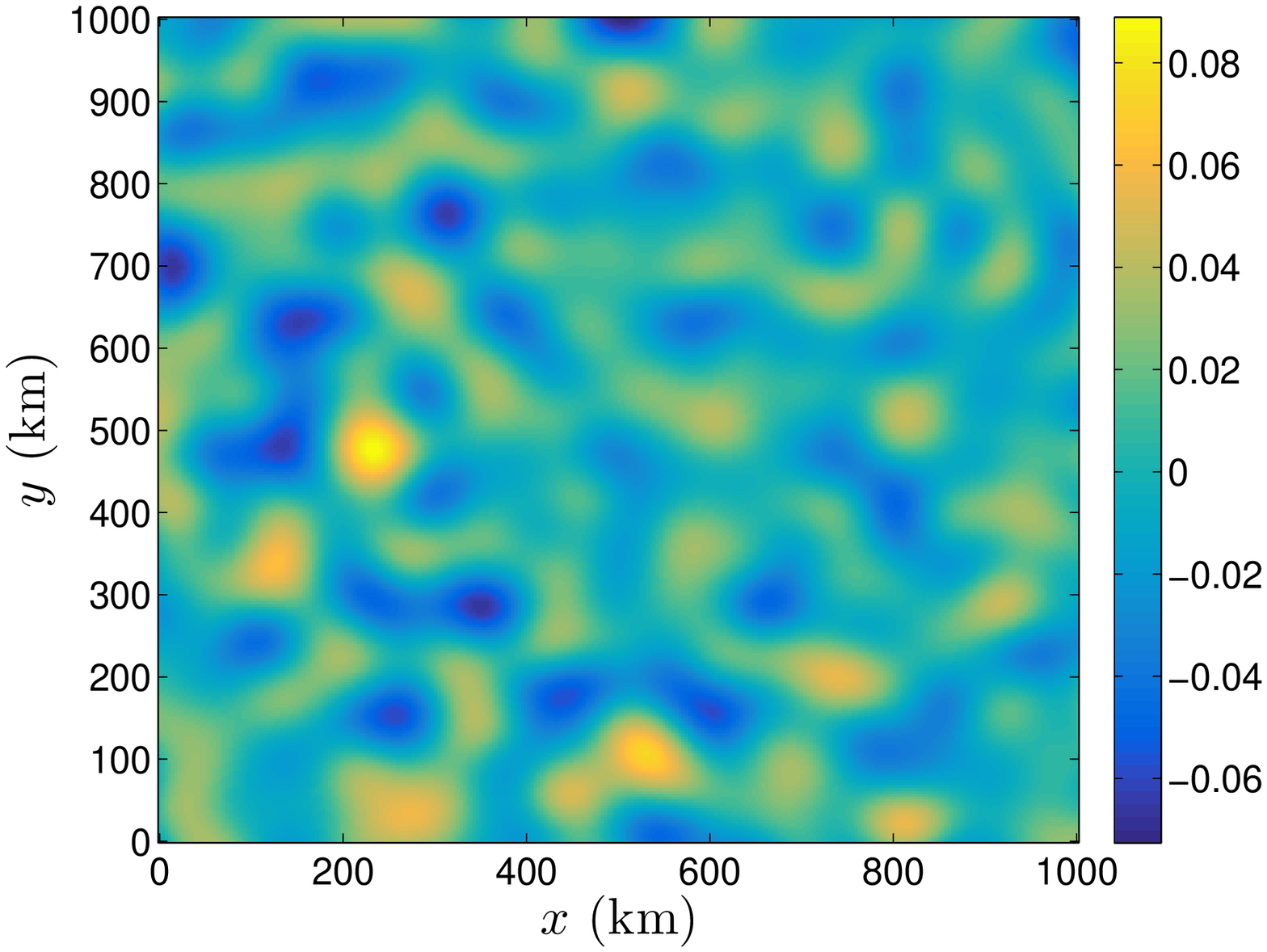}} \\
(c)\raisebox{-.9\height}{\includegraphics[width=.45\textwidth]{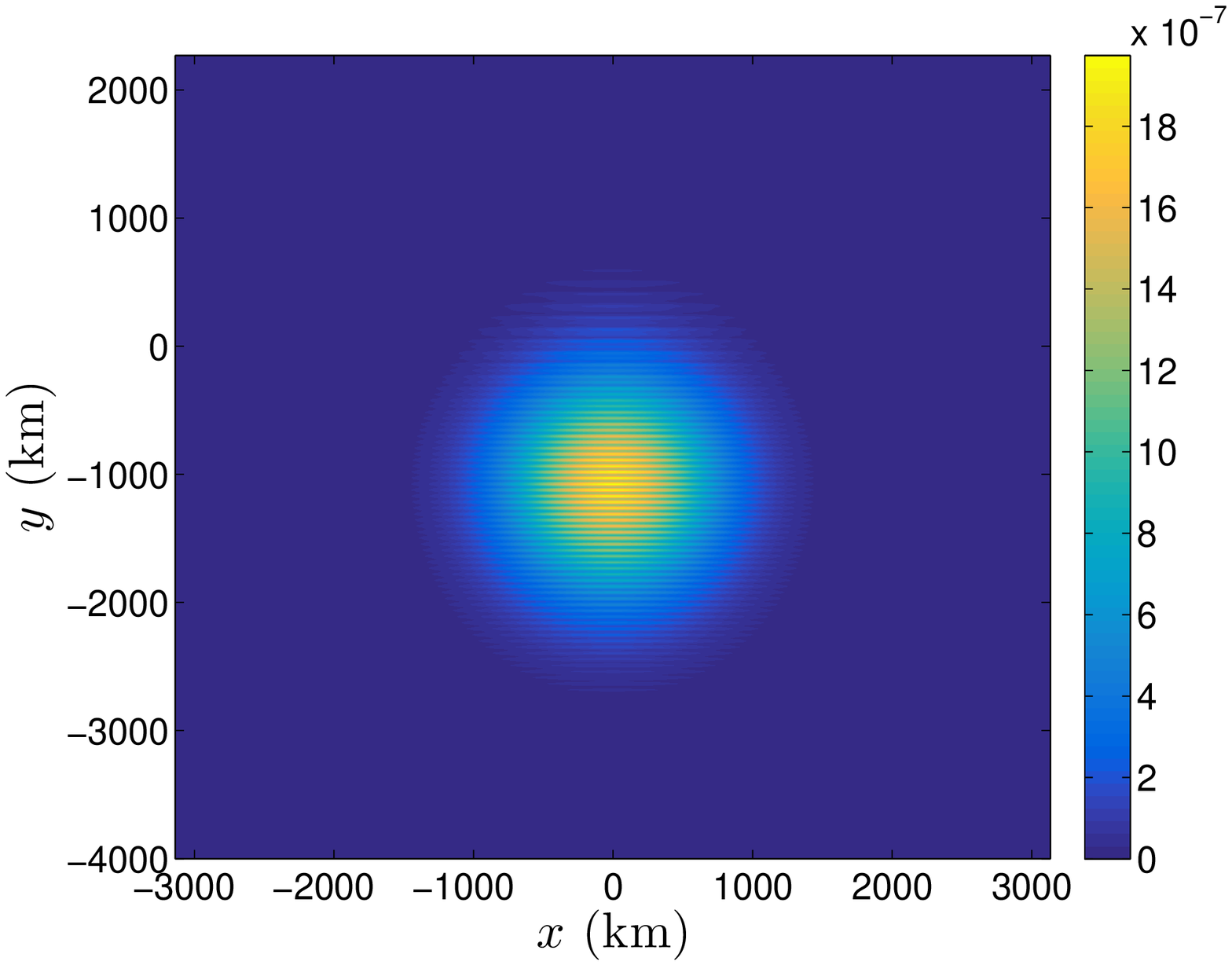}}&
(d)\raisebox{-.9\height}{\includegraphics[width=.45\textwidth]{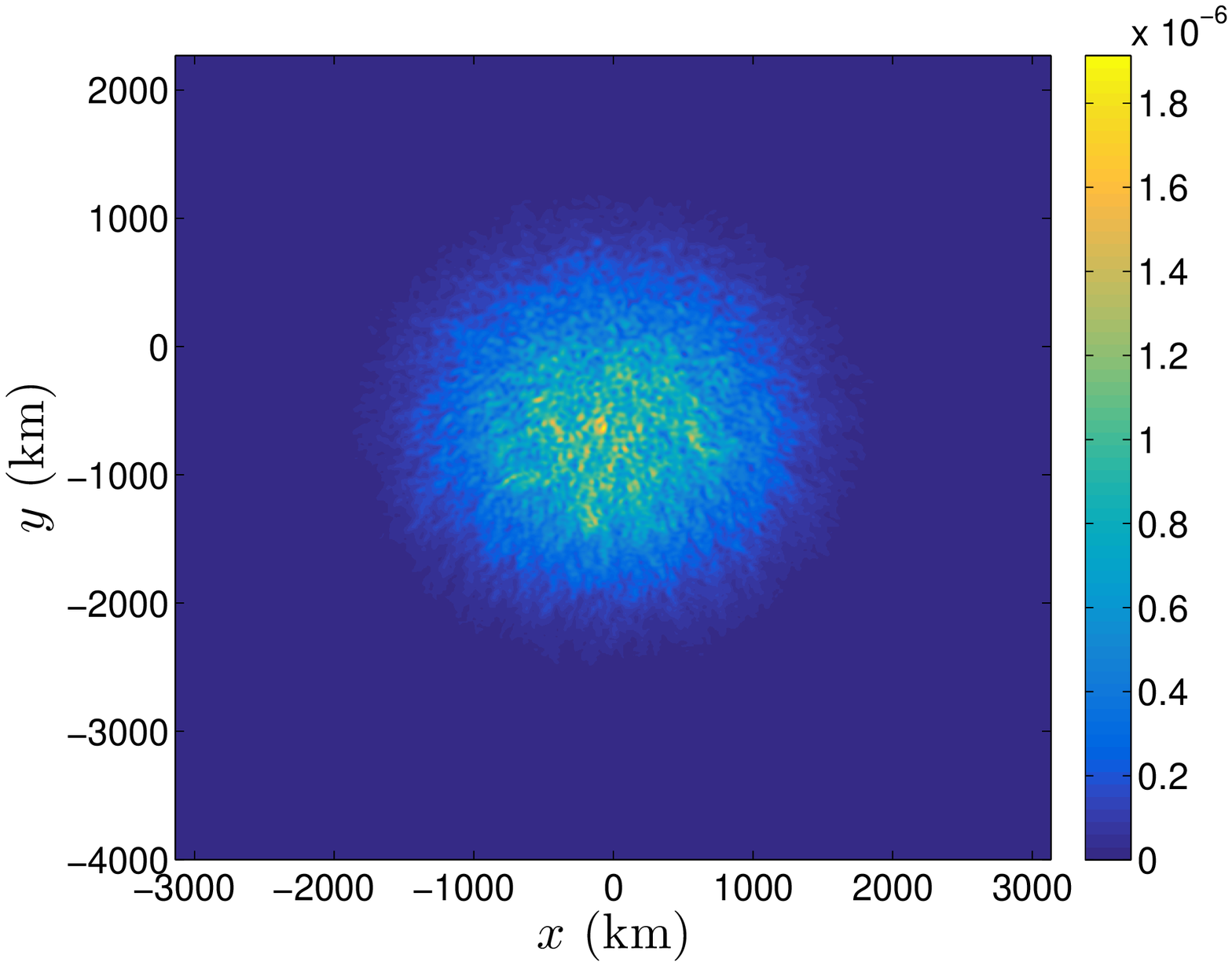}}
\end{tabular}
\caption{Wavepacket on a $\beta$-plane: (a) zonal velocity at the start of the simulations; (b) example of the relative vorticity field (scaled by the Coriolis frequency $f_0$) used in one of the twenty runs, shown in a subdomain of size $1000\times1000$ km; (c) inertial kinetic energy after $30$ days for the simulation with no background flow; (d)  inertial kinetic energy after $30$ days averaged over $20$ flow realisations.}
\label{fig:inertial_wave_beta_plane}
\end{center}
\end{figure}

We examine the propagation of a NIW packet in a region of strong geostrophic turbulence.
Taking the $\beta$-plane approximation $f=f_0+\beta y$ with $f_0=1.16\cdot10^{-4}\text{s}^{-1}$ and $\beta=1.37\cdot10^{-11} \, \text{m}^{-1}\text{s}^{-1}$ (corresponding to a latitude of $53^{\circ}$N), we carry out simulations of an unforced problem, with initial conditions
$$(u,v,\eta)(x,y,t=0)=(\cos(l_0 y),\omega \sin(l_0y)/f_0,Hl_0 \sin(l_0y)/f_0)u_0 \, \e^{-|\bx|^2/(2L^2)},$$
that represent a Gaussian wavepacket satisfying the polarisation relations of inertia-gravity waves. Here, $(u,v)$ is the velocity field, $\eta$ is the perturbation of the mixed-layer depth from its average value $H=50$ m,  $l_0=-2\pi/(180 \, \text{km})$ is the initial meridional wavenumber, and $\omega=\sqrt{f_0^2+g'Hl_0^2}\simeq1.21\cdot10^{-4}\, \text{s}^{-1}$, with $g'=0.02 \, \text{m}\,\text{s}^{-2}$, obeys the inertia-gravity-wave dispersion relation. The other parameters are $L=710$ km and $u_0=0.2 \, \text{cm} \, \text{s}^{-1}$, a value small enough to ensure that the wave--wave interaction terms remain small; it could be increased without affecting much the overall dynamics.  Fig.\ \ref{fig:inertial_wave_beta_plane}a displays the initial zonal wave velocity $u(x,y,t=0)$.

We compare the wave dynamics in a reference simulation without flow and in an ensemble of 20 simulations with flows taken as realisations of a (time-independent) homogeneous and isotropic Gaussian random process with Gaussian correlation function. The r.m.s.\ of the vorticity field in these flows is $\zeta_{\mathrm{rms}}=2.6\cdot10^{-6}\text{s}^{-1}$, and the streamfunction correlation length is $310$ km. A $1000\times1000$ km close-up of the vorticity field in one flow realisation is displayed in Fig.\ \ref{fig:inertial_wave_beta_plane}b. 

\begin{figure}
\begin{center}
\begin{tabular}{cc}
(a)\raisebox{-.9\height}{\includegraphics[width=.45\textwidth]{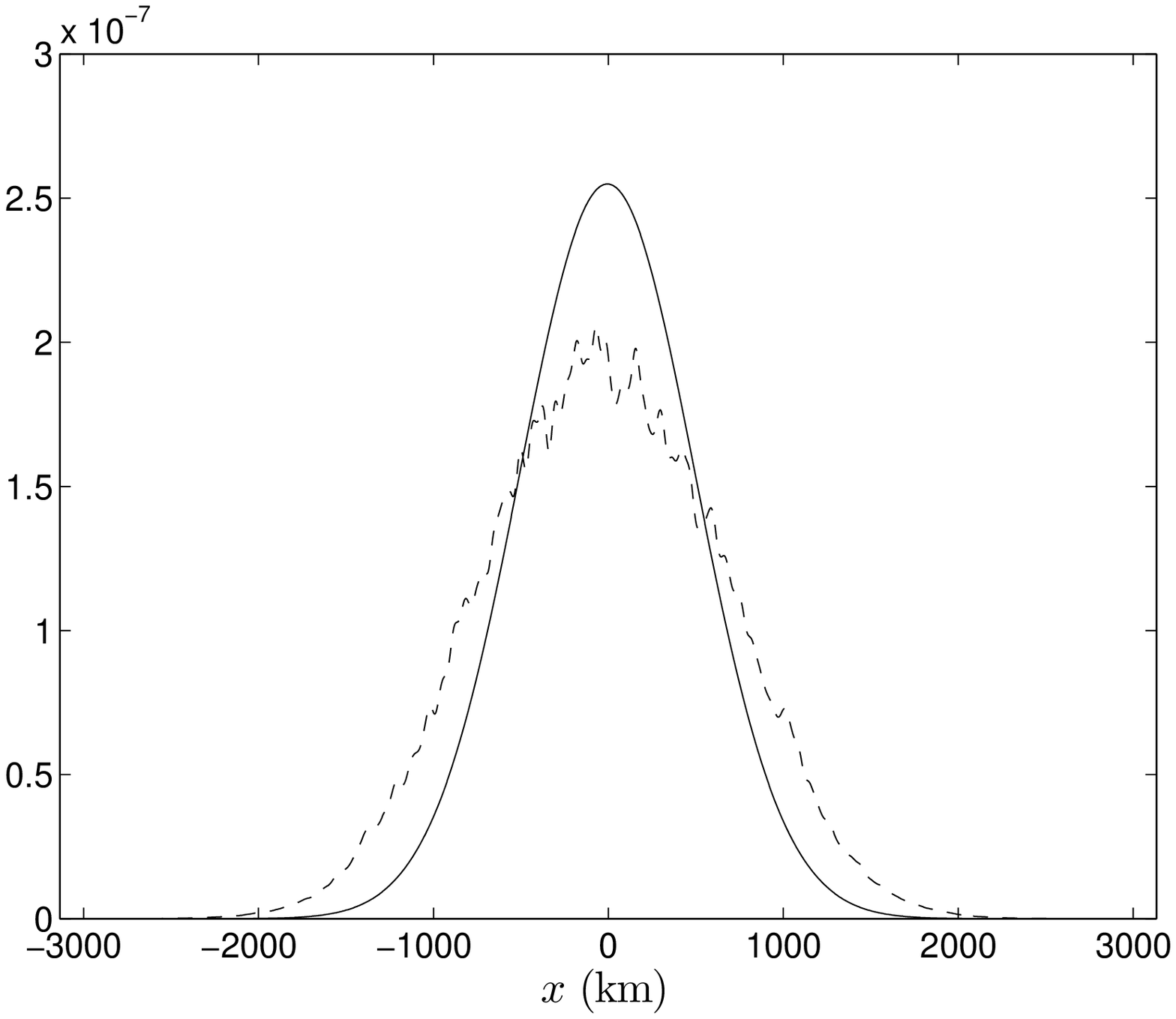}}&
(b)\raisebox{-.9\height}{\includegraphics[width=.45\textwidth]{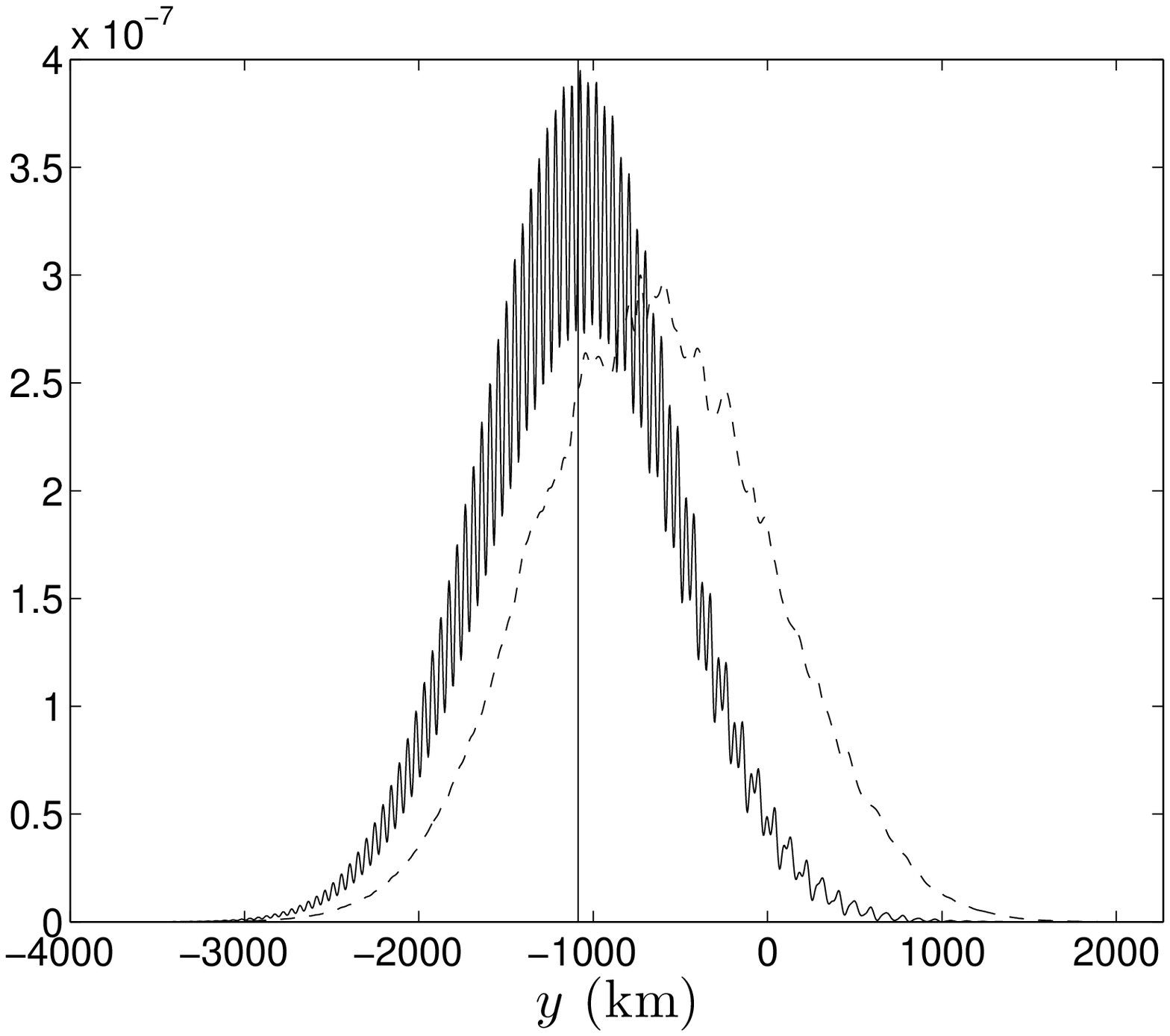}}
\end{tabular}
\caption{Horizontal kinetic energy of the wavepacket on a $\beta$-plane: (a) meridional average in the simulation with no flow (solid line, see Fig.\ \ref{fig:inertial_wave_beta_plane}c) and in the $20$ simulations with flow (dashed line, see Fig.\ \ref{fig:inertial_wave_beta_plane}d), (b) same as (a) but with a zonal average instead. The black vertical line in (b) indicates the location of the wavepacket as predicted by WKB theory.}
\label{fig:inertial_wave_beta_plane_averages}
\end{center}
\end{figure}

Fig.\ \ref{fig:inertial_wave_beta_plane}c illustrates the wavepacket behaviour in the absence of flow by showing the kinetic energy density after 30 days. As expected, the wavepacket drifts equatorward, by a distance $\Delta y\simeq-1100$ km that can be estimated using ray tracing, while the  wavenumber  decreases to $-6.7\cdot10^{-5}\text{m}^{-1}$ to satisfy the dispersion relation. The picture changes considerably in the presence of a flow. Fig.\ \ref{fig:inertial_wave_beta_plane}d shows the kinetic energy density averaged over the $20$ flow realisations; clearly, the wavepacket drifts more slowly equatorward while spreading out more.
Fig.\ \ref{fig:inertial_wave_beta_plane_averages} shows the zonal (panel a) and meridional (panel b) integrals of the  kinetic energy density in Fig.\ \ref{fig:inertial_wave_beta_plane}. The location of the wavepacket calculated from  ray tracing is indicated by a vertical line in  panel b and matches  closely the numerical simulation without flow. The slowdown of the meridional drift and the increase spread in the presence of the flow are evident. To quantify them  we define the drift $\bar{\bx}$ and spread $\sigma$ of the wave kinetic energy distribution $E(\bx)$ from its first and second moments,
$$\bar{\bx}=\frac{\int \bx E(\bx) \, \d\bx}{\int E(\bx)\, \d\bx} \quad 
\textrm{ and } \quad 
\sigma^2=\frac{\int |\bx-\bar{\bx}|^2 E(\bx) \, \d\bx}{\int E(\bx)\,\d\bx}.
$$
After $30$ days, the average meridional drift in the simulations with flow is $\bar y\simeq -650\text{km}$, much smaller than without flow.
The spread with flow is about $\sigma=950$ km, larger than its initial value $710$ km and the final value without flow, about $750$ km. Results are  more dramatic for longer times, but our neglect of vertical propagation then becomes problematic.

The results detailed above are consistent with the scattering process described earlier in the paper.
Let us first note that the $\beta$-effect can be added to the YBJ equation (\ref{eq:YBJ}) and leads to the extra term $\i \beta y M$ on the left-hand side. This modifies the transport equation (\ref{eq:transport}) by adding a transport in spectral space with velocity $-\beta$ in the $l$-direction,
\begin{equation}\label{eq:transport_beta}
\partial_tW+h\bk\cdot\boldsymbol{\nabla}_{\bx} W-\beta \partial_l W=\mathcal{L}W-\Sigma(\bk)W,
\end{equation}
where $\bk=(k,l)$. For brevity, we do not provide a derivation but refer the reader to the general form of the transport equation for a  dispersion relation that depends on both $\bk$ and $\bx$ (Ref.\ \citenum{Ryzhik96}, Eq.\ (1.1)). In the absence of a flow, the transport in spectral space is equivalent to the evolution of the meridional wavenumber $l(t)=l_0-\beta t$ in the WKB approximation.

\begin{figure}
\begin{center}
\begin{tabular}{cc}
(a)\raisebox{-.95\height}{\includegraphics[width=.45\textwidth]{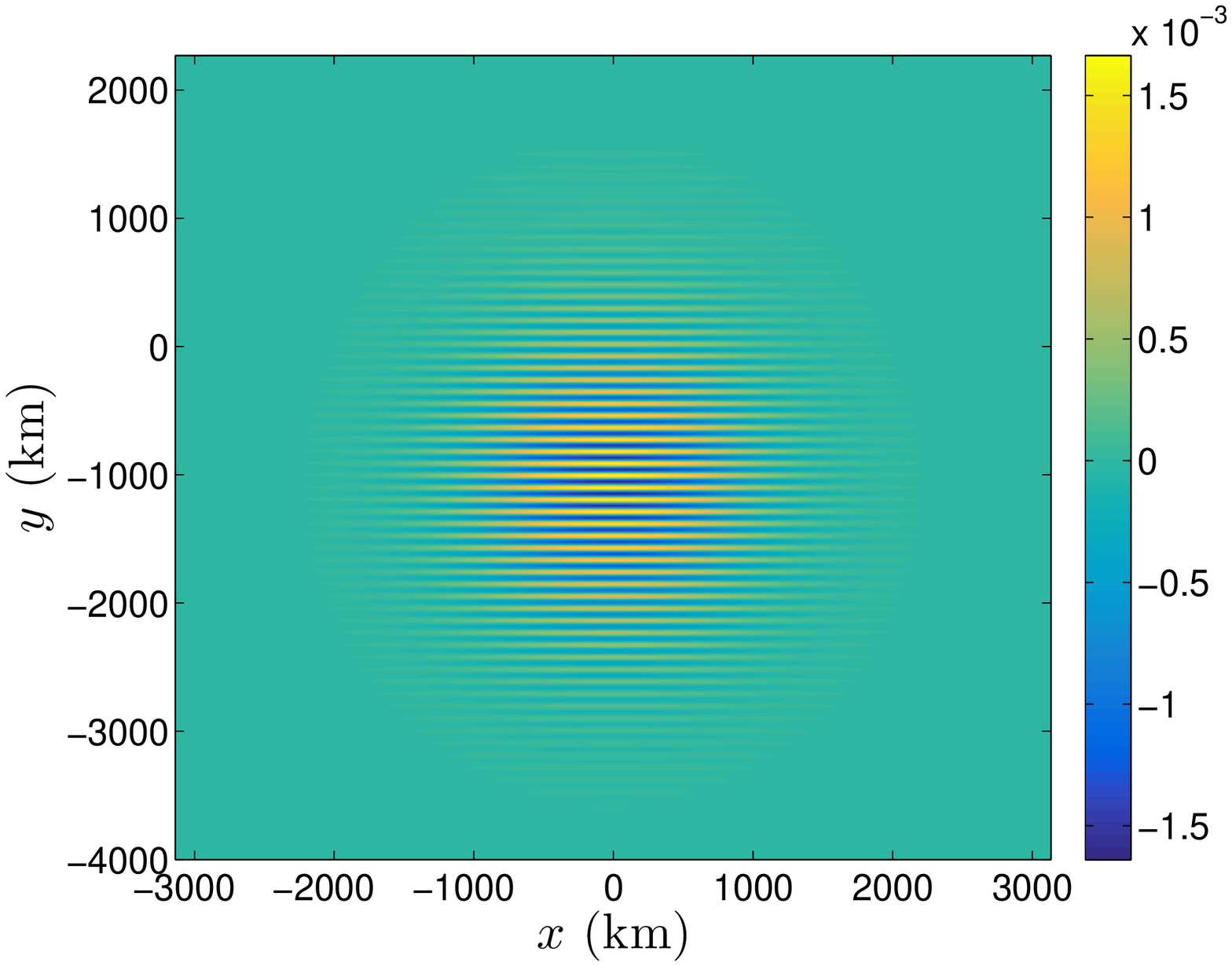}}&
(b)\raisebox{-.95\height}{\includegraphics[width=.45\textwidth]{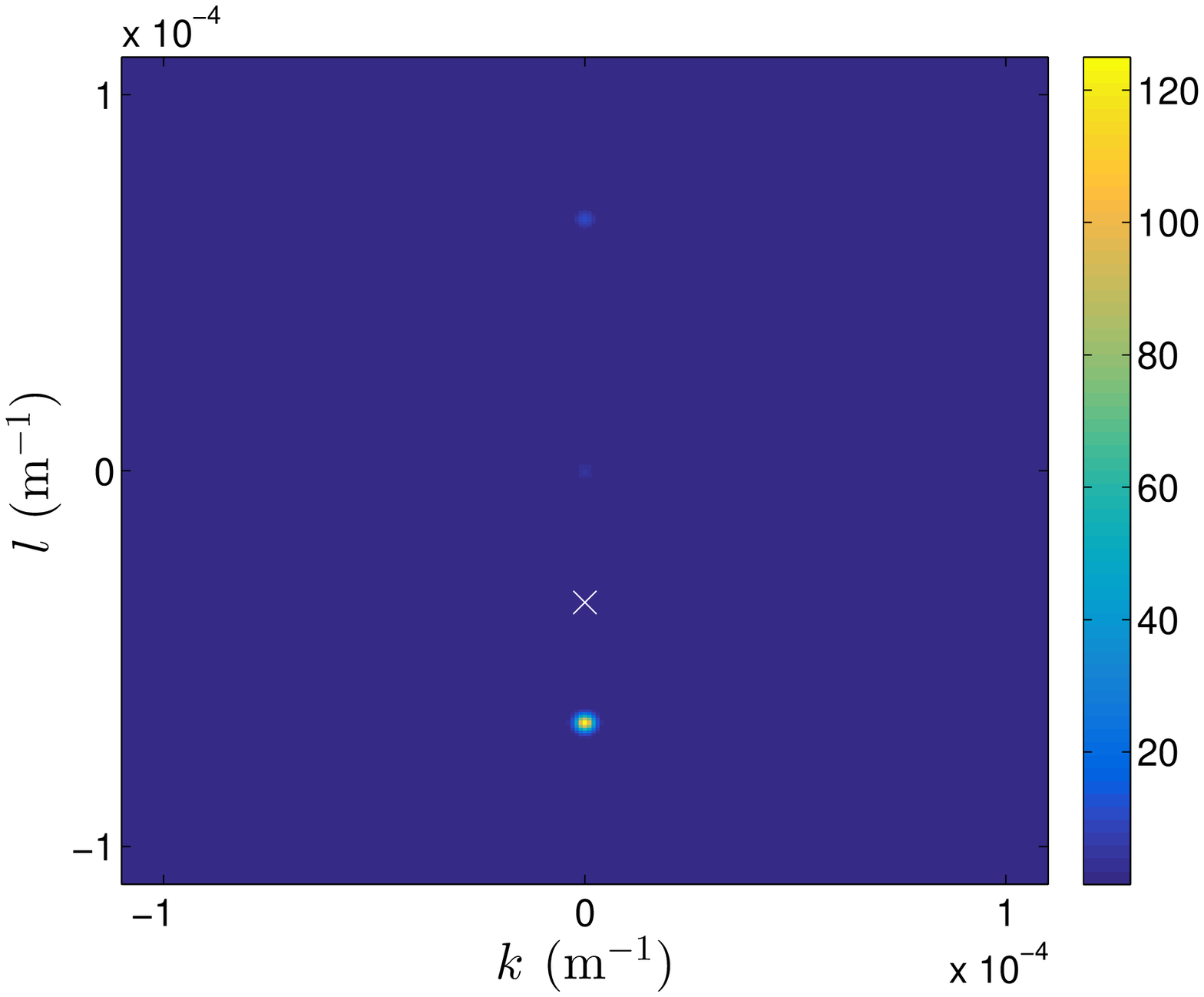}}\\
(c)\raisebox{-.95\height}{\includegraphics[width=.45\textwidth]{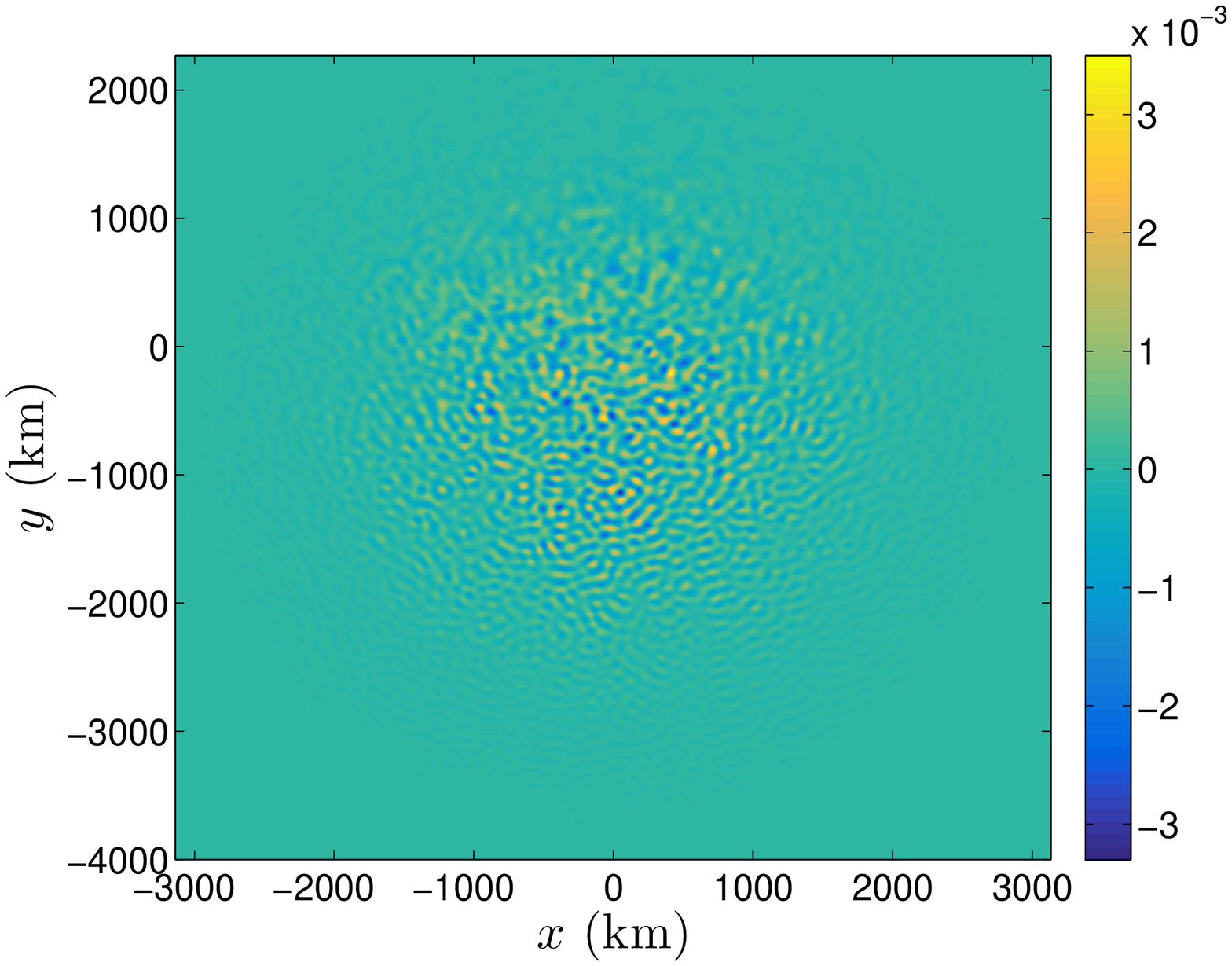}}&
(d)\raisebox{-.95\height}{\includegraphics[width=.45\textwidth]{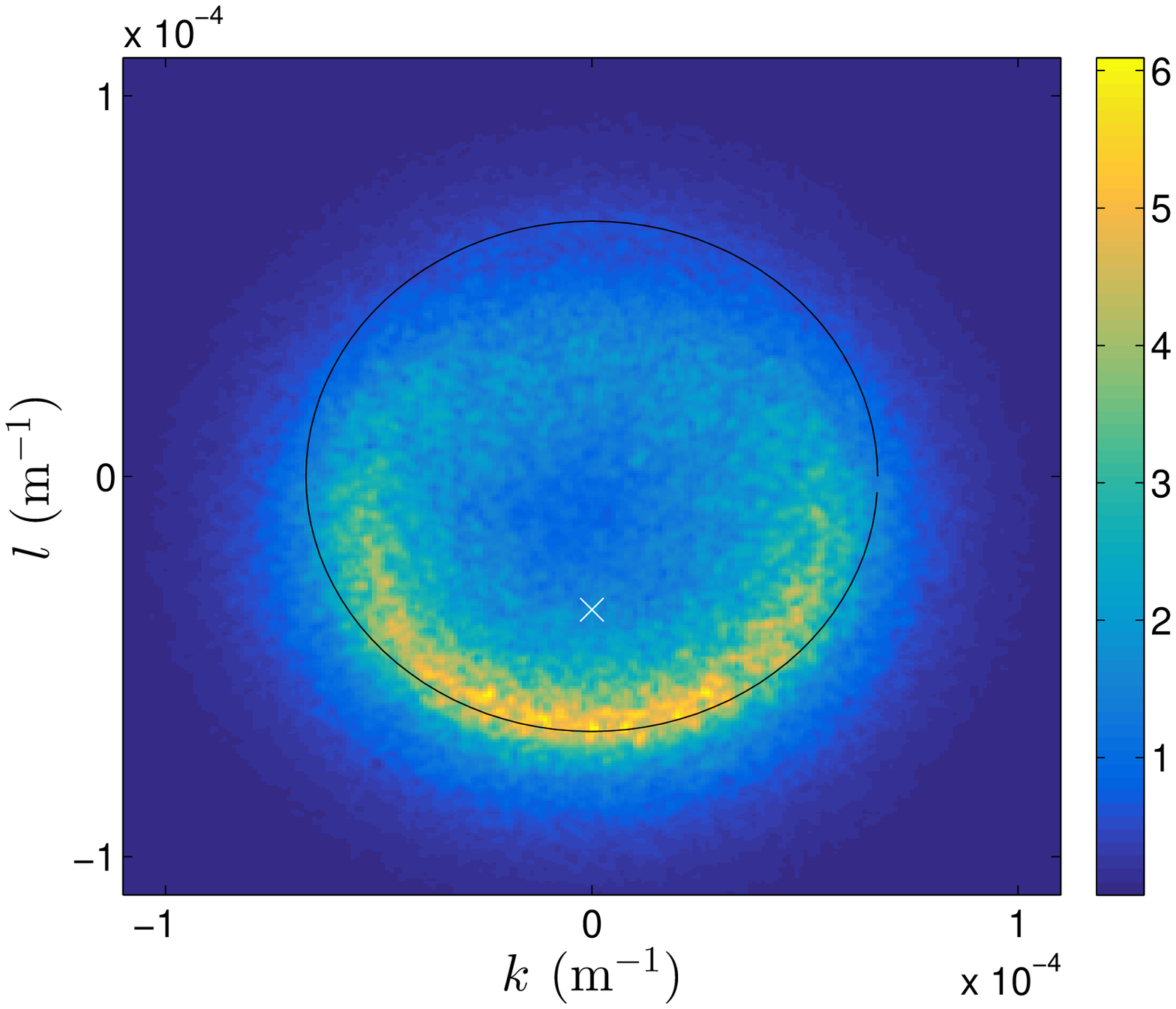}}
\end{tabular}
\caption{Wavepacket  on a $\beta$-plane: zonal velocity (panels a, c) and amplitude of the Fourier transform $|\hat{u}+\i \hat{v}|$ (panels b,d) after $30$ days in the simulations without (a, b) and with flow (c, d). Panel c shows the  zonal velocity in one particular flow realisation whereas panel d shows an average over 20 simulations. Initially the Fourier transform is concentrated in a narrow region centered on $(k,l)=(0,-3.5\cdot10^{-5})$ represented here by a white cross. The circle in panel d has a radius equal to $|l(30\text{days})|=6.7\cdot10^{-5}\text{m}^{-1}$ (see text).
}
\label{fig:phip_phips_beta}
\end{center}
\end{figure}

Qualitatively, (\ref{eq:transport_beta}) implies an interplay between $\beta$-effect and scattering by the flow, the former transferring energy to smaller meridional wavenumbers (larger in magnitude since they are negative), the latter redistributing energy between  wavevectors of the same magnitude.
This interpretation is confirmed by simulations.
Fig.\ \ref{fig:phip_phips_beta} shows the zonal velocity after 30 days without flow (panel a) and in one of the flow realisations (panel c), along with the amplitude of the Fourier transform  of $u + \i v$ (equivalent to $|\hat{M}|$ for NIWs), without flow (panel b), and averaged over the 20 simulations in the case with flow (panel d). Fig.\ \ref{fig:phip_phips_beta}d shows that, as  energy is transferred to smaller meridional wavenumbers due to the $\beta$-effect, it is also transferred to other wavenumbers in a quasi-isotropic way. Because these two processes are concurrent, the energy distribution in spectral space is much more complex than the annulus obtained without $\beta$ (Fig.\ \ref{fig:time_evolution}h). Its shape is close to a circular arc with an energy maximum at $l(t)=l_0-\beta t$ and with a radius of curvature smaller than $|l(t)|$, reflecting the scattering history. Note that the reason why neighbouring wavevectors are excited first, in contrast with the simulation of Fig.\ \ref{fig:time_evolution} is the larger value of $|l(t)|/\kc$, where $\kc$ is related to the flow correlation length through $\kc=2\sqrt{2\pi}/\lc$. Here, $\gamma=2(l(t)/\kc)^2$ is close to 34 after $30$ days (i.e. waves have much smaller scales than the flow), a value much larger than the value $1.1$ used for Fig.\ \ref{fig:time_evolution}. 

As expected from the spectrum in Fig.\ \ref{fig:phip_phips_beta}d, the zonal velocity field (Fig.\ \ref{fig:phip_phips_beta}c) displays a variety of wavelengths, with short waves at the front of the wavepacket and longer waves at the back, while the local wavevector is deflected from its initial North-South orientation into a broad range of directions, causing the spreading of the wavepacket.
We can verify that the time scales associated with the $\beta$-effect and with scattering are comparable:
the former is $l_0/\beta=29$ days; the latter can be estimated by taking the wavenumber as $l(t)=l_0-\beta t$ to find $\Sigma^{-1}=5$ days and $(\Sigma-\lambda')^{-1}=15$ days for $t=0$, and $\Sigma^{-1}=3$ days and $(\Sigma-\lambda')^{-1}=32$ days for $t=30$ days.

\subsection{Moving cyclone}

A moving cyclone generates a wake of internal waves which is stationary in the cyclone reference frame, much like the surface-wave wake generated by a moving ship. Due to resonance, the most energetic waves to emerge are NIWs with a wavenumber $k\sim f_0/U$  in the direction of translation of the cyclone. Here, $f_0$ and $U$ are the Coriolis frequency and cyclone translation speed. A typical value for $U$ is $5$ $\text{m}\,\text{s}^{-1}$, giving a wavelength $2\pi U/f_0\sim300$ km at mid-latitudes, close to the most energetic scales in the ocean \citep{Letraon2008}. Moving cyclones have been studied in detail both in the linear \citep{Geisler70,Nilsson95,Zervakis95} and nonlinear contexts \citep{Chang78,Greatbatch83,Greatbatch84,Price94,Niwa97,Zedler09}. Importantly, all these studies neglect the influence of a background flow by considering an ocean initially at rest.
Therefore, there is a clear need to analyse the propagation of cyclone-generated NIWs in the presence of a turbulent flow. 

Motivated by this, we examine shallow-water simulations of a moving cyclone in $20$ realisations of the random flow used in the previous section ($\zeta_{\mathrm{rms}}=2.6\cdot10^{-6}\, \text{s}^{-1}$, $\lc=310$ km) and, for comparison, one simulation without any flow.
In our simulations, the cyclone is represented by a wind stress of the form
$\boldsymbol{\tau}=\tau_\theta(r)\boldsymbol{e}_\theta$ in a polar coordinates system translating steadily in the eastward direction. The radial profile of the wind stress is
$$
\tau_\theta(r) = \left\{
\begin{array}{ll}
\tau_{\text{max}}{r}/{R}&\textrm{for} \ \ r<R\\
\tau_{\text{max}}(1.2-0.2 {r}/{R})&\textrm{for} \ \  R\leq r\leq 6R\\
0&\textrm{for} \ \ r>6R
\end{array} \right.
$$
(see Ref.\ \citenum{Price83}). The cyclone translation speed is $U=2 \, \text{m} \text{s}^{-1}$ and the radius of maximum stress is $R=75$ km, making this a relatively wide and slow-moving cyclone.
The stress field described above is applied homogeneously throughout the top layer as a body force $\boldsymbol{\tau}/(\rho_0H)$, where $H=100$ m is the average mixed layer depth and $\rho_0=1025 \, \text{kg}\, \text{m}^{-3}$ is a reference density.
To avoid spurious wave emission, $\tau_{\text{max}}$ is progressively increased from zero to the value $\tau_{\text{max}}=0.75\, \text{N}\,\text{m}^{-2}$ (corresponding roughly to 90 km$/$h winds) during the first two days, and damped after $16$ days, before the cyclone, which originates from the western boundary of the domain, hits the eastern boundary. 
The Coriolis frequency is taken as $f_0=7\cdot10^{-5}\, \text{s}^{-1}$, corresponding to a latitude of $30^\circ$N; the $\beta$ effect is neglected for simplicity. Finally, we take $g'=0.01 \, \text{m}\,\text{s}^{-2}$, giving the same short-wave speed $c=(g'H)^{1/2}=1 \, \text{m}\,\text{s}^{-1}$ as in the previous section.

\begin{figure}
\begin{center}
\begin{tabular}{cc}
(a)\raisebox{-.95\height}{\includegraphics[width=.5\textwidth]{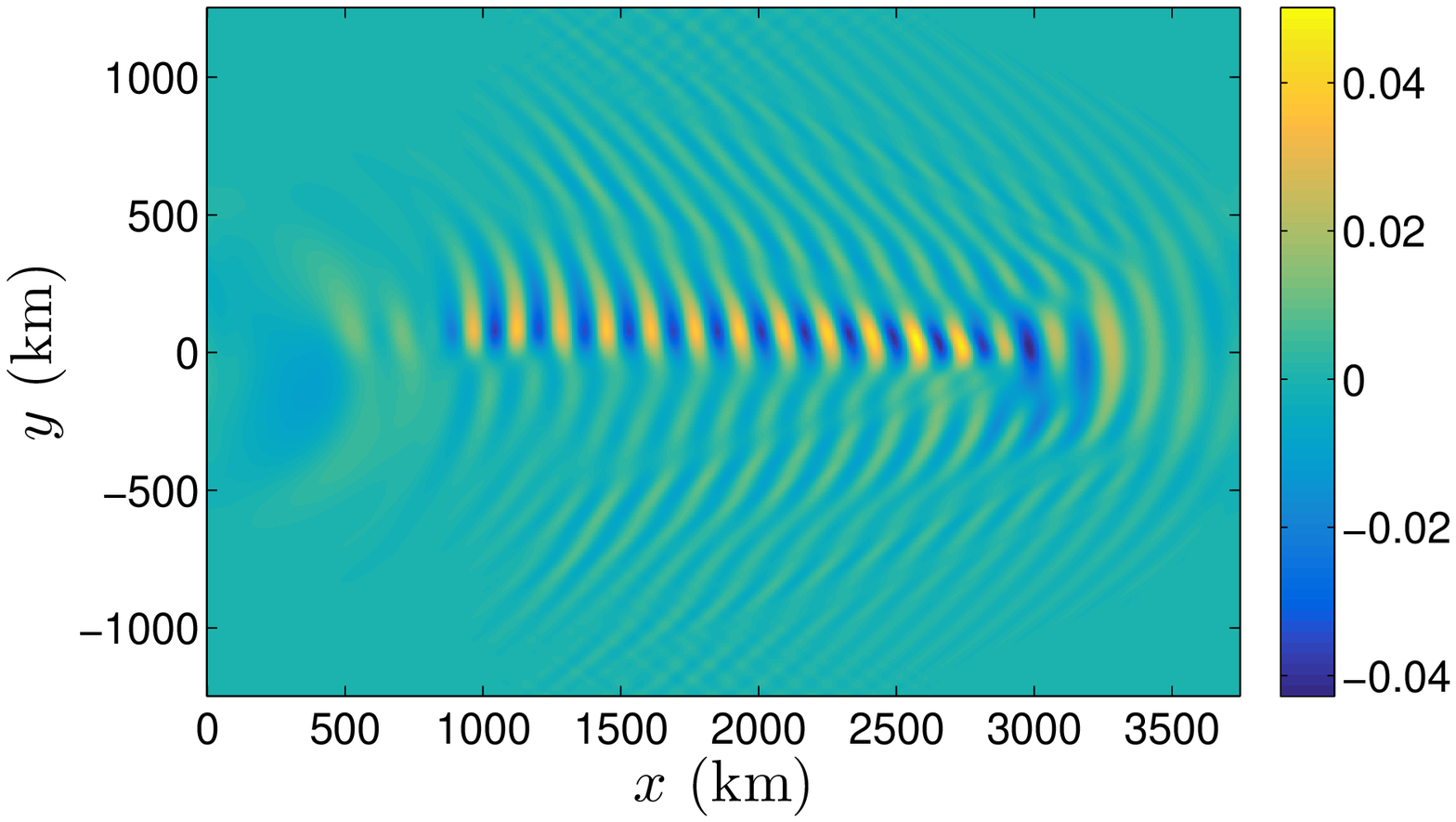}}&
(b)\raisebox{-.95\height}{\includegraphics[width=.4\textwidth]{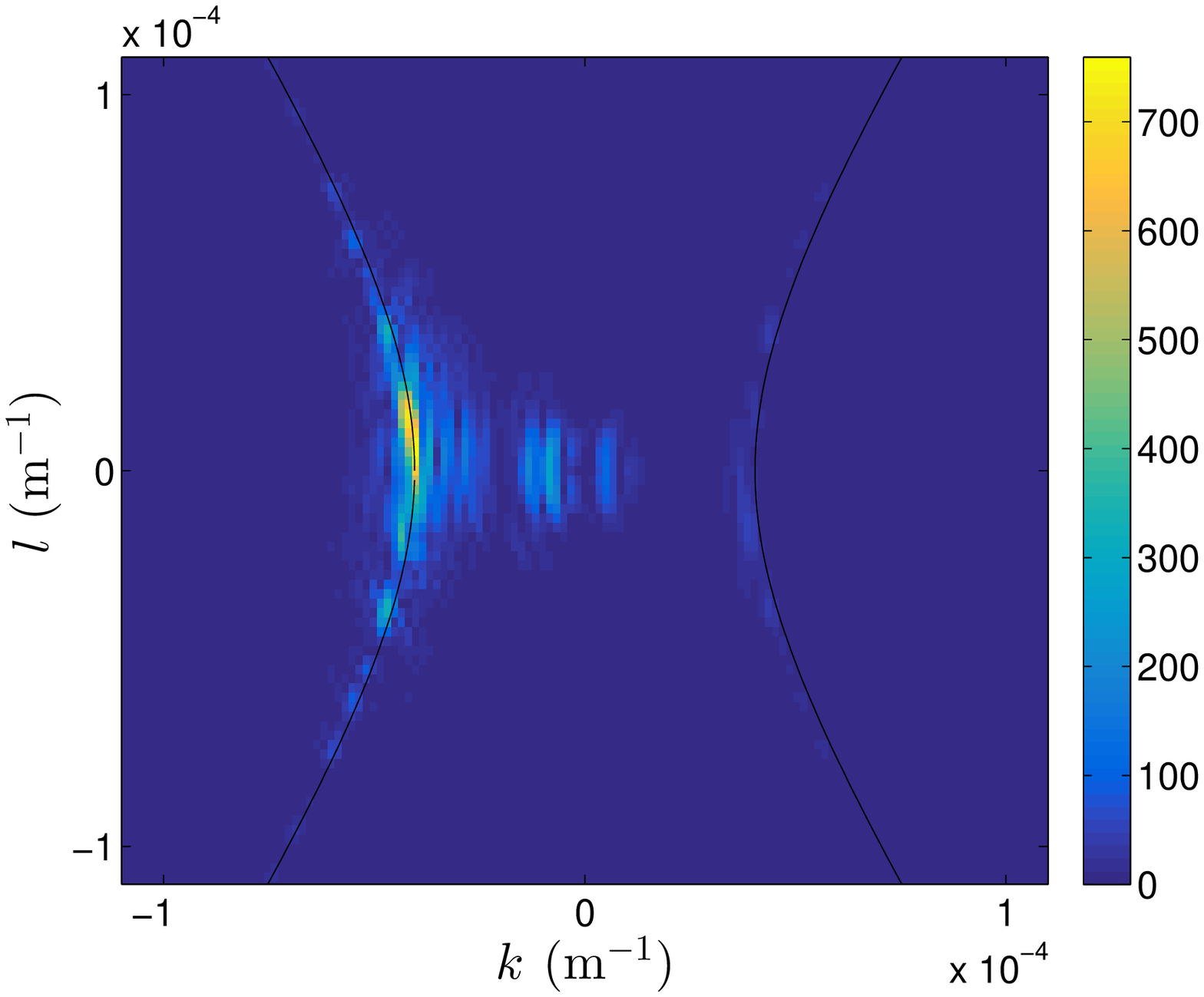}}\\
(c)\raisebox{-.95\height}{\includegraphics[width=.5\textwidth]{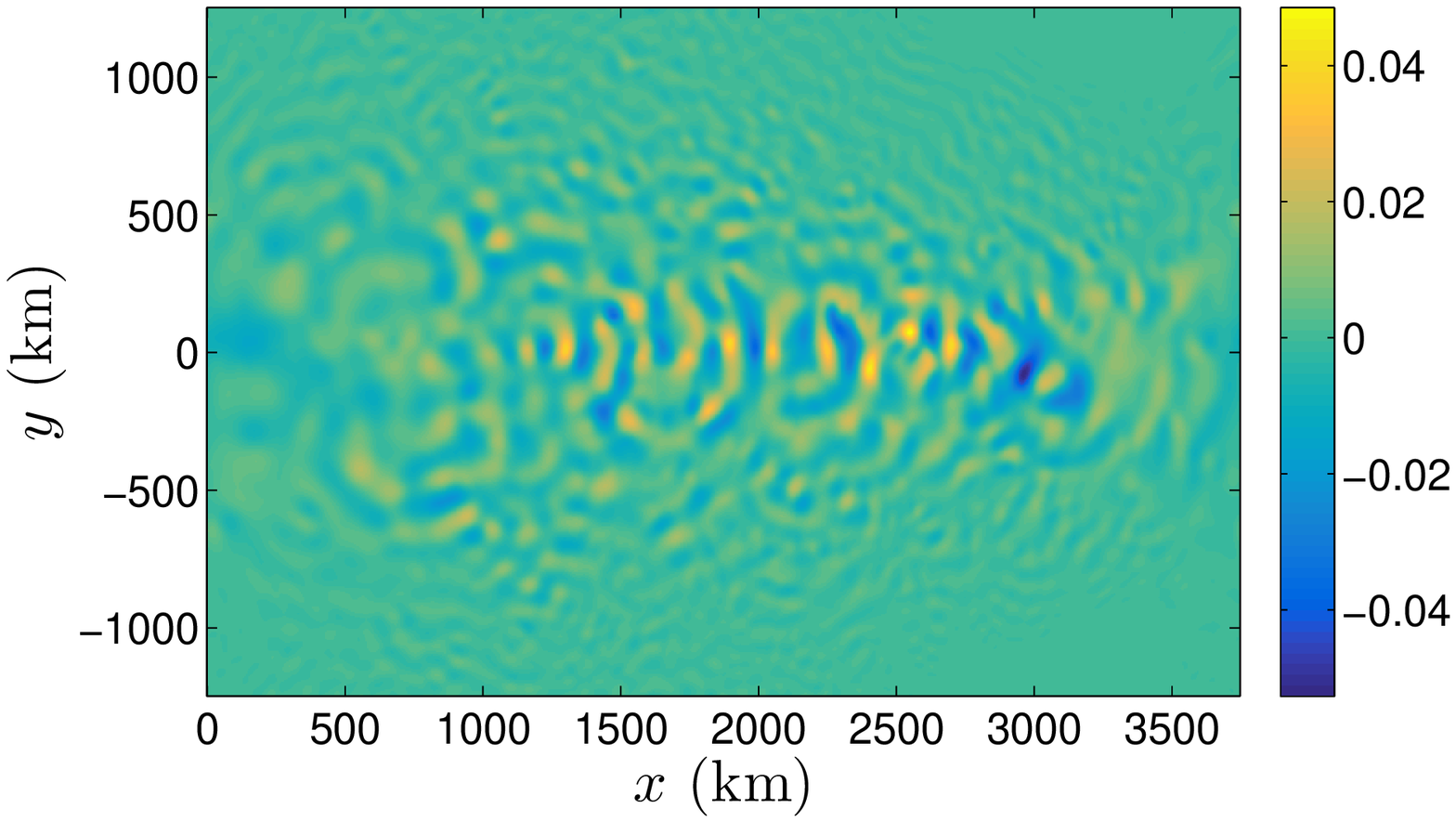}}&
(d)\raisebox{-.95\height}{\includegraphics[width=.4\textwidth]{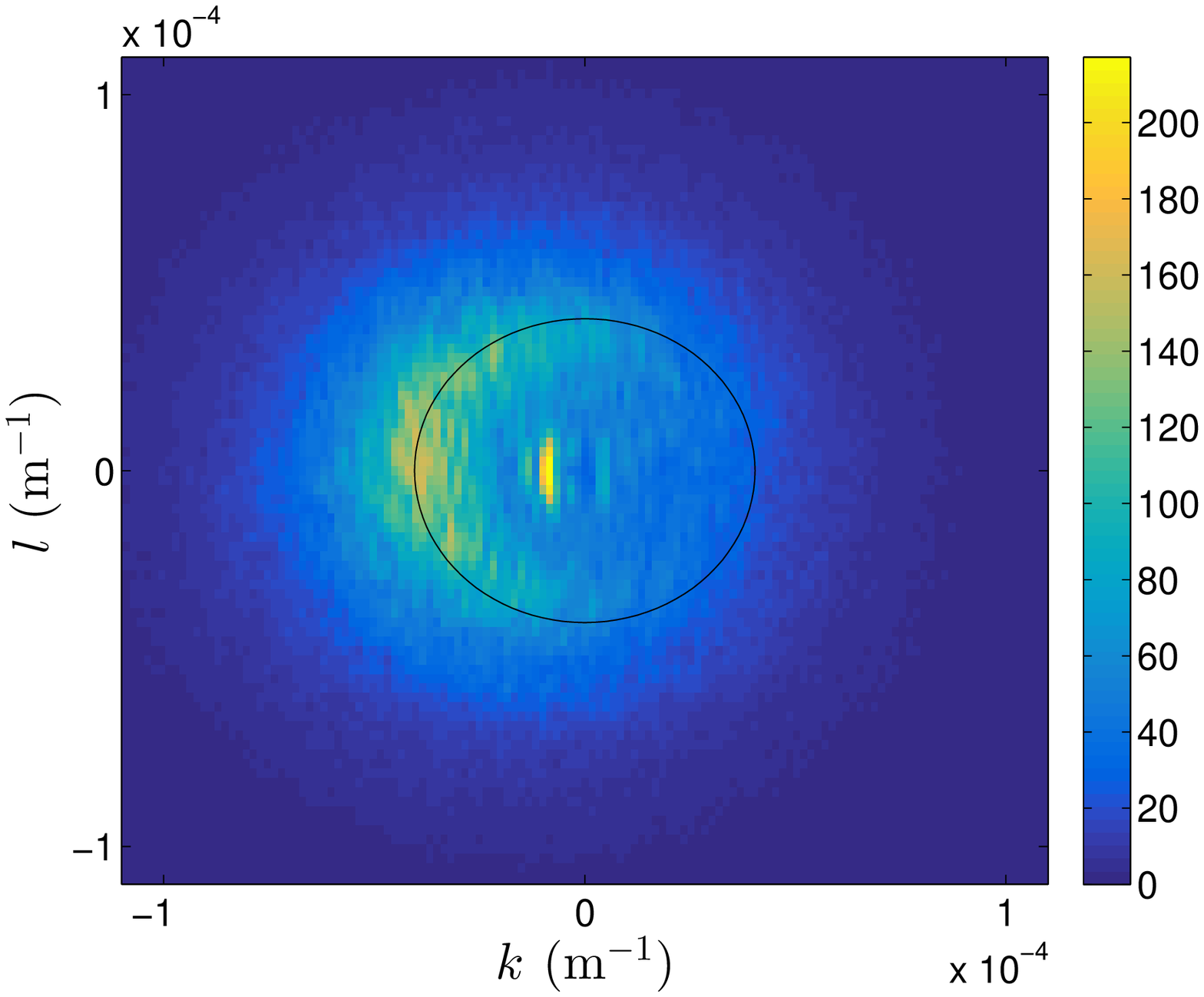}}
\end{tabular}
\caption{NIWs generated by a moving cyclone: near-inertial zonal velocity (panels a, c) and amplitude of the Fourier transform $|\hat{u}+\i \hat{v}|$ (panels b, d) after $25$ days in the simulations without (panels a, b) and with flow (panels c,d). Panel c shows the field in one flow realisation, whereas panel d shows an average over 20 flow realisations. The dispersion relation $(U^2-c^2)k^2=f_0^2+c^2l^2$ and isotropic circle $k^2+l^2=f_0^2/(U^2-c^2)$ are indicated by solid lines in panels b and d, respectively.}
\label{fig:phip_phips_storm}
\end{center}
\end{figure}

Fig.\ \ref{fig:phip_phips_storm} shows the NIW velocity -- extracted from the total velocity field using a time filter -- with and without flow, in physical and spectral spaces, after 25 days. 
Without flow, the wave field generated by a moving cyclone is approximately stationary in the cyclone frame (panel a). The weak North-South asymmetry arises from nonlinear effects, which become significant for wind stresses larger than $\tau_{\text{max}}=0.75 \, \text{N}\,\text{m}^{-2}$.
As time progresses, NIWs spread laterally and the energy at $y=0$ is significantly reduced. Fig.\ \ref{fig:phip_phips_storm}b show that the waves follow the dispersion relation for stationary waves in the cyclone reference frame,  $(Uk)^2=f_0^2+c^2(k^2+l^2)$ (solid lines, see Ref.\ \citenum{Nilsson95}); the energy is concentrated on one branch of the dispersion relation, which can be explained by the anticyclonic rotation rate of NIWs. Note that the energy is not distributed evenly along the dispersion curve: it is  maximum at $(k,l)\sim(k_G,0)$,  where $k_G=f_0/(U^2-c^2)^{1/2}$ is the Geisler wavenumber \citep{Geisler70} (for $c\ll U$, $k_G\sim f_0/U$). Some energy is also present at smaller wavenumbers, possibly due to filtering errors.

\begin{figure}
\begin{center}
\includegraphics[width=.5\textwidth]{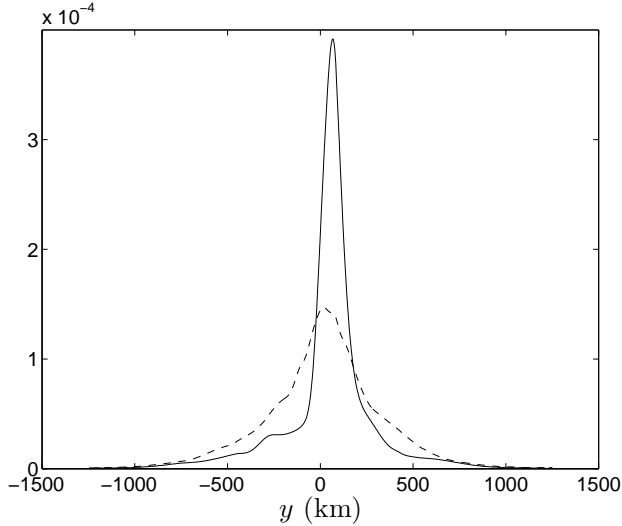}
\caption{NIWs generated by a moving cyclone: zonal average of the NIW horizontal kinetic energy in the simulation with no flow (plain line, see Fig.\ \ref{fig:phip_phips_storm}c) and the $20$ simulations with flow (dashed line) after 25 days.}
\label{fig:KEm_y_moving_cyclone}
\end{center}
\end{figure}

The NIWs are significantly affected by the presence of a flow. In particular, they display small-scale variability in the $y$-direction (Fig.\ \ref{fig:phip_phips_storm}c), with a velocity field that is reminiscent of Figs.\ \ref{fig:time_evolution}g and \ref{fig:phip_phips_beta}c. The magnitude of the Fourier transform $|\hat{u}+\i \hat{v}|$ averaged over the simulations with flow is illuminating (Fig.\ \ref{fig:phip_phips_storm}d): the energy appears to have spread partially along a circle of radius $k_G$, although a maximum is still present at $(k_G,0)$.
The relatively broad structure of energy along the isotropic circle ($k^2+l^2=k_G^2$) stems from nonlinear interactions (not taken into account in the theory): experiences with smaller $\tau_{\text{max}}$ have a more concentrated NIW energy distribution (not shown). 

The results described above can be interpreted in the light of the scattering theory.
The scattering and isotropisation time-scales calculated from the flow characteristics and the Geisler wavenumber are respectively 4 and 23 days, of the order of the simulations length. 
A notable consequence of the (partial) isotropisation process is the acceleration of the lateral dispersion of NIW energy, as illustrated in Fig.\ \ref{fig:KEm_y_moving_cyclone}: the flow deflects waves from their mainly zonal path (Figs.\ \ref{fig:phip_phips_storm}a,b), so that they propagate with a finite angle $\theta=\tan^{-1}(l/k)$ with the $x$-axis (Figs.\ \ref{fig:phip_phips_storm}c,d). As a consequence, energy is transported faster away from the centre of the domain $y=0$.



\section{Discussion} \label{sec:discussion}

In this paper, we examine the impact of a complex, turbulent flow on the propagation of wind-driven near-inertial oceanic waves. We focus on the distinguished regime in which the spatial scale of the flow is comparable to the typical wavelengths. NIWs in this regime are well described by the asymptotic model due to \citet{YBJ} on which our analysis relies. Further simplifications arise from two assumptions made on the flow: first, that its effect is weak compared to dispersion, and second that it can be modelled by a homogeneous stationary random process. With these assumptions, a transport equation governing the energy transfers in physical and spectral spaces is derived, following the general treatment of \citet{Ryzhik96}. This equation describes, in particular, the scattering effect of the flow, which redistributes energy along constant-wavenumber circles in spectral space. For isotropic flows, this ultimately leads to an isotropic wave field, with a scale that is essentially the same as the initial scale. Thus, in the regime considered, dispersion strongly inhibits the cascade to small scales that refraction and advection by the flow can induce.  Two time scales are relevant to the scattering process: the first estimates the time necessary for scattering to be significant, the second the time to achieve isotropisation. Explicit expressions are available for both, with asymptotic expressions showing that they can be very different.  

The theoretical results based on the transport equation derived from the YBJ model are shown to be relevant to the full shallow-water model in two applications. The first is motivated by the observed equatorward propagation of NIWs due to the $\beta$-effect, the second by the generation of mesoscale NIWs in the wake of moving cyclones. In both cases, the isotropisation caused by scattering strongly affects the wave field.

While our work is focussed on NIWs, it can be placed in the more general context of wave--potential-vorticity (PV) interaction. In particular, it is related to work by \citet{Lelong91}, \citet{Bartello95} and \citet{Ward2010} who consider the  triadic interactions between the normal modes of the linear equations, namely internal (or inertia-gravity) waves and the PV (or vortical) mode.
One of the four triadic interactions identified involves a wave-wave-PV resonant triad in which the PV mode, although unaffected, acts as a catalyst for  energy transfers between two indentical-frequency waves. 
This is precisely the interaction that our transport equation captures. We note that this transport equation is derived under the assumption of a given  flow, that is, a (possibly time-dependent) PV mode determined a priori; the weakly nonlinear approach of \citep{Lelong91,Bartello95,Ward2010} makes no such assumption but nonetheless leads to nearly constant PV modes as a result of PV conservation.

The most important feature of the interactions between wave modes catalysed by the flow is that it leads to energy exchanges between waves with the same frequency -- circles in Fourier space for the shallow-water model of the present paper and of \citep{Ward2010}, cones for the continuously stratified models of \citep{Lelong91,Bartello95}. Our assumption of a flow well modelled by a homogeneous random process makes it possible to quantify these exchanges (see \citet{Muller92} for the derivation of a transport equation relevant in the presence of a random topography). One possible extension beyond the near-inertial regime could be to internal tides, following from the recent simulations of \citet{Ponte2015} which show the emergence of spatially complex tidal signal in the presence of a turbulent flow. 

We finally note that our conclusion of a relaxation to an isotropic stationary wave-energy distribution -- corresponding to a uniform energy along constant-frequency curves (circles) in Fourier space -- applies to the shallow-water model but not to the continuously stratified models: for these, the constant-frequency surfaces (cones) are non-compact, and it is therefore impossible for a finite wave energy to relax to a uniform distribution. It would be desirable to obtain the transport equation corresponding to these models and study the associated scattering. We leave this for future work.

\medskip

\noindent
\textbf{Acknowledgements.} This research is funded by the UK
Natural Environment Research Council (grant NE/J022012/1).

\appendix
\section{Derivation of the transport equation}\label{app:deriv}
In this Appendix, we derive  a transport equation for the Wigner function associated with NIWs following \citep{Ryzhik96}. 
This is achieved by assuming a separation between the (small) scale of variation of the phase of the NIWs and the (large) scale of variation of their envelope. The ratio of these two scales is the small parameter $\eps\ll 1$. 
Another key assumption is that the background flow varies on the same scale as the NIW phase.

\subsection{Scaled YBJ equation and Wigner function} 
Because we assume that the NIW amplitude varies on the large scale $L$, it is clear from (\ref{eq:int_Wig}) that the leading-order Wigner function  depends on the large scale only. Therefore, it is natural to work with the dimensionless spatial variable $\bx' = \bx / L$.
We emphasise that this is just a convenient change of coordinate which does not imply that $M$ varies on the large scale; indeed with this choice, $|\boldsymbol{\nabla}_{\bx'}M|=O(\eps^{-1})$. We assume the scaling  $\Psi/h=O(\eps^{1/2})$ for the streamfunction of the background flow and, correspondingly, introduce the dimensionless streamfunction $\psi'=\eps^{-1/2} \psi / h$. This varies over the short spatial scale $\ell$ and thus should be regarded as a function of $\bxi = \eps^{-1} \bx' = \bx/\ell$:  $\psi=\psi(\bxi,t)$. Introducing the non-dimensional time $t' = h t/(\eps L^2)$, we rewrite (\ref{eq:YBJ}) in non-dimensional form
\begin{equation}\label{eq:YBJ2}
\partial_t M + \eps^{1/2}\boldsymbol{\nabla}_{\bxi}^\bot\psi\cdot\boldsymbol{\nabla}_{\bx} M - \eps\frac{\i}{2}\Delta_{\bx} M + \i\eps^{-1/2}\frac{\Delta_{\bxi} \psi}{2} M=0,
\end{equation}
where primes have been omitted and the streamfunction is assumed to vary over the same slow time scale $\eps L^2/ h=L \ell /h $ as the Wigner function.
The correct scaling for the Wigner function in the scale-separation regime is 
\begin{equation}\label{eq:Wig_rescaled}
W^\eps(\bx,\bk,t)=\eps^{-2}W(\bx,\bk/\eps,t)=\frac{1}{4\pi^2}\displaystyle\int \e^{\i\bk\cdot\by}M(\bx-{\eps\by}/{2},t)M^*(\bx+{\eps\by}/{2},t) \, \d\by.
\end{equation}
Note the interesting dual property
\begin{equation}\label{eq:Wig_dual}
W^\eps(\bx,\bk,t)=\eps^{-2}\int \e^{\i\bp\cdot\bx}\tilde{M}(-\eps^{-1}{\bk}-{\bp}/{2},t)\tilde{M}^*(-\eps^{-1}{\bk}+{\bp}/{2},t) \, \d\bp,
\end{equation}
where 
\begin{equation}\label{eq:Fourier_def}
\tilde{M}(\bp,t)=\frac{1}{4\pi^2}\int \e^{\i\bp\cdot\bx}M(\bx,t) \, \d\bx
\end{equation}
is the Fourier transform of $M$ with respect to $\bx$ at wavevector $\bp$.

\subsection{Evolution equation for the Wigner function}
Differentiating (\ref{eq:Wig_rescaled}) with
respect to $t$ and using (\ref{eq:YBJ2}) yields, after some manipulations involving (\ref{eq:Wig_dual}),
\begin{equation}\label{eq:transport1}
\begin{split}
\partial_tW^\eps&+\frac{\eps^{1/2}}{2}\boldsymbol{\nabla}_{\bx} \cdot\int \hat{\bv}(\bl)\e^{-\i\eps^{-1}\bl\cdot\bx}[W^\eps(\bx,\bk+{\bl}/{2})+W^\eps(\bx,\bk-{\bl}/{2})] \, \d\bl\\
&+i\eps^{-{1/2}}\bk\cdot\int \hat{\bv}(\bl)\e^{-\i\eps^{-1}\bl\cdot\bx}[W^\eps(\bx,\bk+{\bl}/{2})-W^\eps(\bx,\bk-{\bl}/{2})] \, \d\bl\\
&+\bk\cdot\boldsymbol{\nabla}_{\bx} W^\eps\\
&+\i\frac{\eps^{{-1/2}}}{2}\int -|\bl|^2\hat{\psi}(\bl)\e^{-\i\eps^{-1}\bl\cdot\bx}[W^\eps(\bx,\bk+{\bl}/{2})-W^\eps(\bx,\bk-{\bl}/{2})] \, \d\bl=0,
\end{split}
\end{equation}
where $\hat{\bv}$ is the Fourier transform of the velocity $\bv(\boldsymbol{\xi},t)\equiv\boldsymbol{\nabla}_\xi^\bot\psi$, i.e.
\begin{equation}\label{eq:Fourier}
\hat{\bv}(\bl)=\frac{1}{4\pi^2}\int \e^{\i\bl\cdot\boldsymbol{\xi}}\bv(\boldsymbol{\xi},t) \, \d\boldsymbol{\xi},
\end{equation}
and similarly $\hat{\psi}$ is the Fourier transform of $\psi$ with respect to $\boldsymbol{\xi}$.
The dependence of $W^\eps$, $\hat{\bv}$ and $\hat \psi$ on time has been omitted for clarity. We stress that the streamfunction, hence the velocity, depend on the slow time variable $t$, not the fast time $t/\eps$ associated with the wave frequency. This assumption justifies  the expansion (\ref{eq:expansion_W}) below. 

Eq.\ (\ref{eq:transport1}) can be rewritten as
\begin{equation}\label{eq:transport2}
\begin{split}
\partial_tW^\eps&+\bk\cdot\boldsymbol{\nabla}_{\bx} W^\eps+\eps^{-1/2}\mathcal{L}^\eps W^\eps\\
&+\frac{\eps^{1/2}}{2}\int \hat{\bv}(\bl)\e^{-\i\eps^{-1}\bl\cdot\bx}\cdot\boldsymbol{\nabla}_{\bx} [W^\eps(\bx,\bk+{\bl}/{2})+W^\eps(\bx,\bk-{\bl}/{2})]\, \d\bl = 0,
\end{split}
\end{equation}
where
$$
\mathcal{L}^\eps W^\eps=\i\int \hat{V}(\bk,\bl)\e^{-\i\eps^{-1}\bl\cdot\bx}[W^\eps(\bx,\bk+{\bl}/{2})-W^\eps(\bx,\bk-{\bl}/{2})]\, \d\bl
$$
is the sum of the third and fifth terms of equation (\ref{eq:transport1}). The potential $V$ is defined through its Fourier transform 
\begin{equation}\label{eq:defV}
\hat{V}(\bk,\bl)=\bk\cdot\hat{\bv}(\bl)-{|\bl|^2}\hat{\psi}(\bl)/2=(i\bk\times\bl-{|\bl|^2}/{2})\hat{\psi}(\bl),
\end{equation}
with $\bk\times\bl=k_1l_2-k_2l_1$ for $\bk=(k_1,k_2)$ and  $\bl=(l_1,l_2)$. Note that (\ref{eq:transport2}) makes use of the non-divergence of the background flow velocity $\boldsymbol{\nabla}_{\boldsymbol{\xi}} \cdot\bv=0$.

\subsection{Asymptotic expansion, random flow and transport equation}
We can now derive a transport equation from (\ref{eq:transport2}) using a multiscale approach that treats $\bx$ and $\boldsymbol{\xi}$ as independent variables,: expanding $W^\eps$ in powers of $\eps^{1/2}$ and assuming that the leading order does not depend on the small scale, we write
\begin{equation}\label{eq:expansion_W}
W^\eps(\bx,\bxi,\bk,t)=W^{(0)}(\bx,\bk,t)+\eps^{1/2}W^{(1)}(\bx,\bxi,\bk,t)+\eps W^{(2)}(\bx,\bxi,\bk,t)+O(\eps^{3/2}).
\end{equation}

At order $\eps^{-1/2}$, we find the same balance as in Ref.\ \citenum{Ryzhik96},
\begin{equation}\label{eq:W1a}
\bk\cdot\boldsymbol{\nabla}_{\bxi} W^{(1)}+\theta W^{(1)}=-\mathcal{L}^\eps W^{(0)},
\end{equation}
where $\theta$ is a regularization parameter which will be set to zero later. The solution is  easily obtained in Fourier space as
\begin{equation}\label{eq:W1b}
\hat{W}^{(1)}(\bx,\bp,\bk,t)=\hat{V}(\bk,\bp)\hat{Y}(\bx,\bp,\bk,t),
\end{equation}
where $\hat{W}^{(1)}$ is the Fourier transform of $W^{(1)}$ with respect to $\boldsymbol{\xi}$ and
\begin{equation}\label{eq:Y}
\hat{Y}(\bx,\bp,\bk,t)=\frac{W^{(0)}(\bx,\bk+{\bp}/{2},t)-W^{(0)}(\bx,\bk-{\bp}/{2},t)}{\bk\cdot\bp+\i\theta}.
\end{equation}

At the next order, $O(\eps^0)$, we find
\begin{multline}\label{eq:W0a}
\partial_tW^{(0)}+\bk\cdot\boldsymbol{\nabla}_{\bx} W^{(0)}
+\bk\cdot\boldsymbol{\nabla}_{\bxi} W^{(2)}+\mathcal{L}^\eps W^{(1)}\\+\frac{1}{2}\int \e^{-\i\eps^{-1}\bl\cdot\bx} \hat{\bv}(\bl)\cdot\boldsymbol{\nabla}_{\bxi}[W^{(1)}(\bx,\bxi,\bk+{\bl}/{2},t)+W^{(1)}(\bx,\bxi,\bk-{\bl}/{2},t)] \, \d\bl=0
\end{multline}

We now introduce the statistical average $\la\cdot\ra$. It can be thought as an ensemble average or, equivalently, as an average over $\bxi$.
Assuming that the flow is homogeneous, we define the covariance $R$ of the streamfunction as
\begin{equation}\label{eq:cov}
R(\boldsymbol{\xi}-\boldsymbol{\xi'})=\la \psi(\boldsymbol{\xi})\psi(\boldsymbol{\xi'})\ra.
\end{equation}
Its Fourier transform with respect to $\boldsymbol{\xi}$, the  power spectrum, is related to $\hat{\psi}$ through
\begin{equation}\label{eq:cov_spec}
\la\hat{\psi}(\bl)\hat{\psi}(\bl')\ra=\hat{R}(\bl)\delta(\bl+\bl'),
\end{equation}
where $\delta$ is the Dirac distribution.

Averaging (\ref{eq:W0a}) and assuming that the small-scale average of the Wigner function is supported by $W^{(0)}$, i.e. $\la W^\eps\ra=\la W^{(0)}\ra$, leads to
\begin{multline}\label{eq:W0b}
\partial_tW^{(0)}+\bk\cdot\boldsymbol{\nabla}_{\bx} W^{(0)}
+\la\mathcal{L}^\eps W^{(1)}\ra
\\
+\la\,\,\frac{1}{2}\int \e^{-\i {\eps}^{-1} \bl\cdot{\bx}} \hat{\bv}(\bl)\cdot\boldsymbol{\nabla}_\xi[W^{(1)}(\bx,\xi,\bk+{\bl}/{2},t)+W^{(1)}(\bx,\xi,\bk-{\bl}/{2},t)] \, \d\bl\,\,\ra=0,
\end{multline}
since $\la\boldsymbol{\nabla}_{\bxi} W^{(2)}\ra=0$ and $\la W^{(0)}\ra= W^{(0)}$.

Using (\ref{eq:W1b}) and (\ref{eq:Y}), it can be shown that the last term of (\ref{eq:W0b}) is equal to
\begin{multline*}
\frac{1}{2}\iint \e^{-\i\eps^{-1}(\bl+\bl')\cdot \bx}
\bl'\times\bl\,[i(\bk+{\bl}/{2})\times\mathbf{l'}-{|\bl'|^2}/{2}]
\la\hat{\psi}(\bl)\hat{\psi}(\bl')\ra\hat{Y}(\bx,\bl',\bk+{\bl}/{2},t) \, \d\bl \d\bl'\\
+
\frac{1}{2}\iint \e^{-\i\eps^{-1}(\bl+\bl')\cdot \bx}
\bl'\times\bl\,[\i(\bk-{\bl}/{2})\times\bl'-{|\bl'|^2}/{2}]
\la\hat{\psi}(\bl)\hat{\psi}(\bl')\ra\hat{Y}(\bx,\bl',\bk-{\bl}/{2},t) \, \d\bl\d\bl',
\end{multline*}
which, from (\ref{eq:cov_spec}),  clearly vanishes.
Hence the transport equation for $W^{(0)}$ reduces to
\begin{equation}\label{eq:transport_final}
\partial_tW^{(0)}+\bk\cdot\boldsymbol{\nabla}_x W^{(0)}=\overline{\mathcal{L}^\eps}W^{(0)},
\end{equation}
where 
\begin{equation}\label{eq:scat_coef}
\overline{\mathcal{L}^\eps}W^{(0)}=4\pi\int[|\bk\times\bp|^2+{|\bp-\bk|^4}/{4}]\hat{R}(\bp-\bk)\delta(\bk^2-\bp^2)\left(W^{(0)}(\bx,\bp,t)-W^{(0)}(\bx,\bk,t)\right)\, \d\bp
\end{equation}
is obtained from $\la\mathcal{L}^\eps W^{(1)}\ra$ in (\ref{eq:W0b}) using (\ref{eq:W1b}), (\ref{eq:Y}), (\ref{eq:cov_spec}) and setting $\theta\rightarrow 0$ (see Ref.\ \citenum{Ryzhik96}).
Eq.\ (\ref{eq:scat_coef}) involves two terms: the first, proportional to $|\bp-\bk|^4/4$ is due to refraction and is the only one present for the Schr\"odinger equation \citep{Ryzhik96}; the second, proportional to $|\bk\times\bp|^2$, appears as a result of advection.
The total scattering cross-section is 
$$\Sigma=4\pi\int [(\bk\times\bp)^2+{|\bp-\bk|^4}/{4}]\hat{R}(\bp-\bk)\delta(\bk^2-\bp^2)\, \d\bp.$$
Note that (\ref{eq:transport}) is the dimensional version of (\ref{eq:transport_final}). 


\section{Eigenvalues of $\mathcal{L}$ for $|\bk| \gg \kc$}\label{app:steepest}
We approximate the eigenvalues $\lambda_n$ (see (\ref{eq:lambda_n})) for $|\bk|\gg \kc$. Because $\sigma'(|\bk|,\theta)=\sigma'(|\bk|,-\theta)$, $\lambda_n$ is equal to
\begin{equation}\label{eq:lambda_n1}
\lambda_n=\frac{8\pi|\bk|^4A}{h}\int_{-\pi}^\pi\sin^2({\theta}/{2}) \exp(-\gamma\sin^2({\theta}/{2})+in\theta) \, \d\theta,
\end{equation}
where $\gamma=2|\bk|^2/\kc^2$ and we have used $\hat{R}(|\bp|)=A\exp(-|\bp|^2/(2\kc^2))$ in (\ref{eq:lambda_n1}). We describe in \ref{app:steepest1} the steepest-descent method used to approximate (\ref{eq:lambda_n1}) for finite $n/\gamma$;  an approximation for small $n$ is derived in \ref{app:steepest2} by Laplace's method.
\subsection{Uniform approximation}\label{app:steepest1}
We derive an approximation valid uniformly for $n=O(1)$ and $n=O(\gamma) \gg 1$ by writing $n=\gamma m$. Eq.\ (\ref{eq:lambda_n1}) then becomes
\begin{equation}\label{eq:lambda_n2}
\lambda_n=\frac{16\pi|\bk|^4A}{h}\int_{-\pi/2}^{\pi/2}\sin^2\theta\exp(-\gamma g(\theta)) \, \d\theta,
\end{equation}
where $g(\theta)=\sin^2\theta-2\i m\theta$. Because $\gamma\gg 1$ in the regime considered here, the integral is dominated by the contribution of the integrand at the critical points of $g(\theta)$, and a steepest-descent method may be applied.
Solving $g'(\theta)=0$ gives one single critical point 
\begin{equation}\label{eq:theta}
\theta^*=\i/2\sinh^{-1}(2m)\in i\mathbb{R}.
\end{equation}
Because $g''(\theta^*)=2\cos(2\theta^*)>0$, one can change the path of integration in (\ref{eq:lambda_n2}) from the real-axis segment $[-\pi/2,\pi/2]$ to a curve $\mathcal{C}$ in the complex plan with the same end points and passing through $\theta^*$, where it crosses the imaginary axis orthogonally. Eq.\ (\ref{eq:lambda_n2}) then becomes approximately
\begin{equation}\label{eq:lambda_n3}
\begin{split}
\lambda_n\simeq &\frac{16\pi|\bk|^4A}{h}\exp(-\gamma g(\theta^*))\int_\mathcal{C}d\theta\sin^2\theta\\
&\times\exp\left(-\gamma \frac{g''(\theta^*)}{2}(\theta-\theta^*)^2
-\gamma \frac{g^{(3)}(\theta^*)}{6}(\theta-\theta^*)^3
-\gamma \frac{g^{(4)}(\theta^*)}{24}(\theta-\theta^*)^4\right).
\end{split}
\end{equation}

It is important to keep all orders in the exponential up to $(\theta-\theta^*)^4$ in order to calculate the first-order correction to $\lambda_n$. 
Introducing the Taylor expansion
\begin{equation}\label{eq:sin_taylor}
\sin^2\theta\simeq\sin^2\theta^*+\sin(2\theta^*)(\theta-\theta^*)+\cos(2\theta^*)(\theta-\theta^*)^2-\frac{2}{3}\sin(2\theta^*)(\theta-\theta^*)^3-\frac{1}{3}\cos(2\theta^*)(\theta-\theta^*)^4
\end{equation}
into the integral (\ref{eq:lambda_n3}) and Taylor-expanding the exponential gives
\begin{equation}\label{eq:lambda_n4}
\begin{split}
\lambda_n\simeq& \frac{16\pi|\bk|^4A}{h}\exp(-\gamma g(\theta^*))\times\\
&\left(\sin^2\theta^*\int_\mathcal{C}\exp(-\gamma \frac{g''(\theta^*)}{2}(\theta-\theta^*)^2)  \, \d\theta\right.\\
&+\cos(2\theta^*)\int_\mathcal{C}(\theta-\theta^*)^2\exp(-\gamma \frac{g''(\theta^*)}{2}(\theta-\theta^*)^2) \, \d\theta\\
&-\gamma\left(\sin(2\theta^*)\frac{g^{(3)}(\theta^*)}{6}+\sin^2\theta^*\frac{g^{(4)}(\theta^*)}{24}\right)\int_\mathcal{C}(\theta-\theta^*)^4\exp(-\gamma \frac{g''(\theta^*)}{2}(\theta-\theta^*)^2)  \, \d\theta\\
&\left.+\gamma^2\sin^2\theta^*\frac{(g^{(3)}(\theta^*))^2}{72}\int_\mathcal{C}(\theta-\theta^*)^6\exp(-\gamma \frac{g''(\theta^*)}{2}(\theta-\theta^*)^2)  \, \d\theta\right).
\end{split}
\end{equation}
Note that we have discarded some terms in (\ref{eq:lambda_n4}) because they appear at a higher order in $\gamma^{-1}$. Moreover we have used symmetry properties to eliminate some integrals.
The integrals appearing in (\ref{eq:lambda_n4}) are standard and we obtain, for large $\gamma$,
\begin{equation}\label{eq:lambda_n5}
\begin{split}
\lambda_n\simeq &\frac{16\pi|\bk|^4A}{h}\exp(-\gamma g(\theta^*))\times\left(\sin^2\theta^*\sqrt{\frac{2\pi}{ g''(\theta^*)}}\gamma^{-1/2}+\cos(2\theta^*)\frac{\sqrt{2\pi}}{ g''(\theta^*)^{3/2}}\gamma^{-3/2}\right.\\
&-\left.\left(\sin(2\theta^*)\frac{g^{(3)}(\theta^*)}{6}+\sin^2\theta^*\frac{g^{(4)}(\theta^*)}{24}\right)\frac{3\sqrt{2\pi}}{g''(\theta^*)^{5/2}}\gamma^{-3/2}
+\sin^2\theta^*\frac{15\sqrt{2\pi}(g^{(3)}(\theta^*))^2}{72g''(\theta^*)^{7/2}}\gamma^{-3/2}
\right).
\end{split}
\end{equation}
The presence of the last two terms in (\ref{eq:lambda_n5}), at the same order as the second term (proportional to $\cos(2\theta^*)$), justifies a posteriori the Taylor expansion of $g(\theta)$ near $\theta^*$ in the exponential in (\ref{eq:lambda_n3}).
Finally, using the value of $\theta^*$ (\ref{eq:theta}) and some trigonometric identities yields (\ref{eq:lambda_n_formula}).
\subsection{Approximation for small $n/\gamma$}\label{app:steepest2}
Eq.\ (\ref{eq:lambda_n1}) can be rewritten as 
\begin{equation}\label{eq:lambda_n6}
\lambda_n=\frac{16\pi|\bk|^4A}{h}\int_{-\pi/2}^{\pi/2}\cos(2n\theta)\sin^2\theta\exp(-\gamma \sin^2\theta) \, \d\theta,
\end{equation}
For large $\gamma$, this integral is dominated by the contribution of the integrand near $\theta=0$. Therefore, we Taylor-expand there the various terms to obtain
\begin{equation}\label{eq:lambda_n7}
\lambda_n\simeq \frac{16\pi|\bk|^4A}{h}\int_{-\pi/2}^{\pi/2}(1-2n^2\theta^2)(\theta^2-\theta^4/3)(1+{\gamma}\theta^4/3)\exp(-\gamma\theta^2) \, \d\theta.
\end{equation}
Re-arranging terms this becomes
\begin{equation}\label{eq:lambda_n8}
\lambda_n\simeq \frac{16\pi|\bk|^4A}{h}\int_{-\pi/2}^{\pi/2}\left(1-(2n^2+{1}/{3})\theta^2+{\gamma}\theta^4/3\right)\theta^2\exp(-\gamma\theta^2) \, \d\theta,
\end{equation}
and, on integrating,
\begin{equation}\label{eq:lambda_n9}
\lambda_n\simeq \frac{8\pi^{3/2}|\bk|^4A}{h}\gamma^{-3/2}\left(1+\gamma^{-1}({3}/{4}-3n^2)\right),
\end{equation}
which yields (\ref{eq:lambda_n_formula_1}).

\end{document}